\documentclass[conference]{IEEEtran}
\IEEEoverridecommandlockouts
\usepackage{amsmath,amsfonts}
\usepackage{algorithmic}
\usepackage{algorithm}
\usepackage{array}
\usepackage[caption=false,font=scriptsize,labelfont=rm,textfont=rm]{subfig}
\usepackage{textcomp}
\usepackage{stfloats}
\usepackage{url}
\usepackage{verbatim}
\usepackage{float}
\usepackage{graphicx}
\usepackage{cite}
\usepackage{hyperref}
\usepackage{amssymb}

\makeatletter
\newenvironment{breakablealgorithm}
{
		\begin{center}
			\refstepcounter{algorithm}
			\hrule height.8pt depth0pt \kern2pt
			\renewcommand{\caption}[2][\relax]{
				{\raggedright\textbf{\ALG@name~\thealgorithm} ##2\par}%
				\ifx\relax##1\relax 
				\addcontentsline{loa}{algorithm}{\protect\numberline{\thealgorithm}##2}%
				\else 
				\addcontentsline{loa}{algorithm}{\protect\numberline{\thealgorithm}##1}%
				\fi
				\kern2pt\hrule\kern2pt
			}
		}{
		\kern2pt\hrule\relax
	\end{center}
}
\makeatother

\begin{document}

\title{An Effective Index for Truss-based Community Search on Large Directed Graphs}
\author{
	\IEEEauthorblockN{Wei Ai}
	\IEEEauthorblockA{\textit{School of Computer and Information Engineering} \\
		\textit{Central South University of Forestry and Technology}\\
		ChangSha, China \\
		aiwei@hnu.edu.cn}
	\and
	\IEEEauthorblockN{CanHao Xie}
	\IEEEauthorblockA{\textit{School of Computer and Information Engineering} \\
		\textit{Central South University of Forestry and Technology}\\
		ChangSha, China \\
		Xiecanhao@csuft.edu.cn}
	\and
	\IEEEauthorblockN{\hspace{5em}Tao Meng{*}}
	\IEEEauthorblockA{\textit{\hspace{4em}School of Computer and Information Engineering}\\
		\textit{\hspace{4em}Central South University of Forestry and Technology}\\
		\hspace{4em}ChangSha, China \\
		\hspace{4em}mengtao@hnu.edu.cn}
	\and
	\IEEEauthorblockN{Yinghao Wu}
	\IEEEauthorblockA{\textit{School of Computer and Information Engineering} \\
		\textit{Central South University of Forestry and Technology}\\
		ChangSha, China \\
		wuyinghao@csuft.edu.cn}
	\and
	\IEEEauthorblockN{\hspace{7em}KeQin Li}
	\IEEEauthorblockA{\hspace{8em}\textit{Department of Computer Science} \\
		\textit{\hspace{8em}State University of New York}\\
		\hspace{8em}	New Paltz, New York 12561, USA \\
		\hspace{8em}lik@newpaltz.edu}
	\thanks{* is the corresponding author.}
}

\maketitle

\begin{abstract}
	Community search is a derivative of community detection that enables online and personalized discovery of communities and has found extensive applications in massive real-world networks. Recently, there needs to be more focus on the community search issue within directed graphs, even though substantial research has been carried out on undirected graphs. The recently proposed D-truss model has achieved good results in the quality of retrieved communities. However, existing D-truss-based work cannot perform efficient community searches on large graphs because it consumes too many computing resources to retrieve the maximal D-truss. To overcome this issue,  we introduce an innovative merge relation known as D-truss-connected to capture the inherent density and cohesiveness of edges within D-truss. This relation allows us to partition all the edges in the original graph into a series of D-truss-connected classes. Then, we construct a concise and compact index, ConDTruss, based on D-truss-connected. Using ConDTruss, the efficiency of maximum D-truss retrieval will be greatly improved, making it a theoretically optimal approach. Experimental evaluations conducted on large directed graph certificate the effectiveness of our proposed method.
\end{abstract}


\section{Introduction}

\IEEEPARstart{C}{ommunity} structure\cite{girvan2002community, leicht2008community} exists widely in large-scale network graphs. The current research on communities can be divided into two categories; one is community detection\cite{xie2013overlapping, leskovec2010empirical, hric2014community}, which aims to divide the network nodes into several communities to understand the network structure and function better; the other is community search(CS)\cite{sozio2010community, kong2019k, li2015influential, fang2020survey, xu2022efficient}, which supports finding specific communities or groups online. 
More specifically, when given a vertex $q$ in a graph $G_d$, CS aims to find all dense and cohesive subgraphs in $G_d$ that contain the vertex $q$. The exploration of community search has gained significant prominence as it empowers individuals to satisfy their requirements more effectively. Concurrently, it has also witnessed substantial real-world utilization, such as personalized recommendations and advertisements, finding research on a specific field or topic papers. In addition, CS research is not limited to simple graphs, but also explores complex graph types, including temporal graphs\cite{li2018persistent}, geo-social graphs\cite{chen2018maximum, fang2017effectives, zhu2017geo}, attributed graphs\cite{fang2017effective, huang2017attribute}, weighted graphs\cite{li2017most}, and multi-valued graphs\cite{li2018skyline}.

Currently, the models mainly include based on k-clique, k-core, and k-truss\cite{sariyuce2016incremental, huang2014querying, cui2013online}. The previously mentioned models are exclusively applied to undirected graphs. However, directed graphs are prevalent in diverse domains, e.g., sensor networks\cite{kandris2020applications}, knowledge map\cite{zins2007knowledge}, social networks,\cite{brandes2013social} and beyond\cite{kim2010finding, leicht2008community}. Consequently, addressing the identification of communities within large directed graphs is of utmost importance.

Several models have been developed to identify cohesive communities within directed graphs, e.g., D-core (also known as ($k$, $l$)-core)\cite{giatsidis2013d} and CF-truss\cite{fang2018effective}. Nevertheless, the D-core model is characterized by a significant drawback. In certain graphs, there can be substantial variations in the in-degree and out-degree of different vertices \cite{shou2022conversational, shou2022object, ying2021prediction, meng2021multi, shou2023graph, shou2023czl, shou2023comprehensive, ai2023gcn, meng2023deep, shou2023adversarial}. It will lead to sparse communities when attempting to incorporate such nodes. On the other hand, the CF-truss model treats the two types of triangles independently, resulting in huge communities. This approach limits its applicability to real-world queries.

Recently, Huang et al.\cite{liu2020truss} performed CS research on directed graphs with a D-truss(($k_{c}$, $k_{f}$)-truss) model. The D-truss model stands out due to its robust structure and cohesiveness. Specifically, each edge in a D-truss can form cyclic (flow) triangles with at least  $k_c$($k_f$) nodes. Moreover, a D-truss $H_s$ is a maximal D-truss(M-D-truss) when no other D-truss $H_s'$ in the original graph satisfies the condition $H_s'$ $\supset$ $H_s$. 
Based on the D-truss model, they propose the definition of the D-truss CS and prove its NP-hardness. In order to enable D-truss CS to proceed, they first devised two algorithms, e.g., $Local$ and $Global$\cite{liu2020truss}, to identify D-truss communities in a down-top and top-down manner, respectively. Then, they designed the D-truss index to acquire the M-D-truss. In particular, they initially introduced a D-truss decomposition algorithm designed to break down the original graph into a series of D-trusses and store the outcome within an index. The basic idea of index-based M-D-truss finding is to use the query vertex set $Q$ as starting points and employ a breadth-first search to identify all the edges of the M-D-truss. Nevertheless, this approach entails excessive edge access and computations, resulting in markedly inefficient community discovery processes within large-scale graphs. Consider an example in Fig 1. Suppose we want to retrieve a M-D-truss in a graph. In that case, we need to calculate and visit the skyline trussnesses of all edges (skyline trussness will be introduced in Section 3), which is undoubtedly very time-consuming.

\IEEEpubidadjcol
{\bf{Example 1.}} In the directed graph $G_d$ displayed in Fig 1, the subgraph $H$ is a (1,1)-truss composed of nodes $'2'$, $'3'$, $'6'$, and $'7'$ and the edges between them, and each edge in $H$ can form at least one flow triangle and one circle triangle with other nodes. However, $H$ is not an M-(1, 1)-truss because other subgraphs in $G_d$ are also (1, 1)-truss and can contain $H$.

\begin{figure}[H]
	\centering
	\includegraphics[width=3.5in]{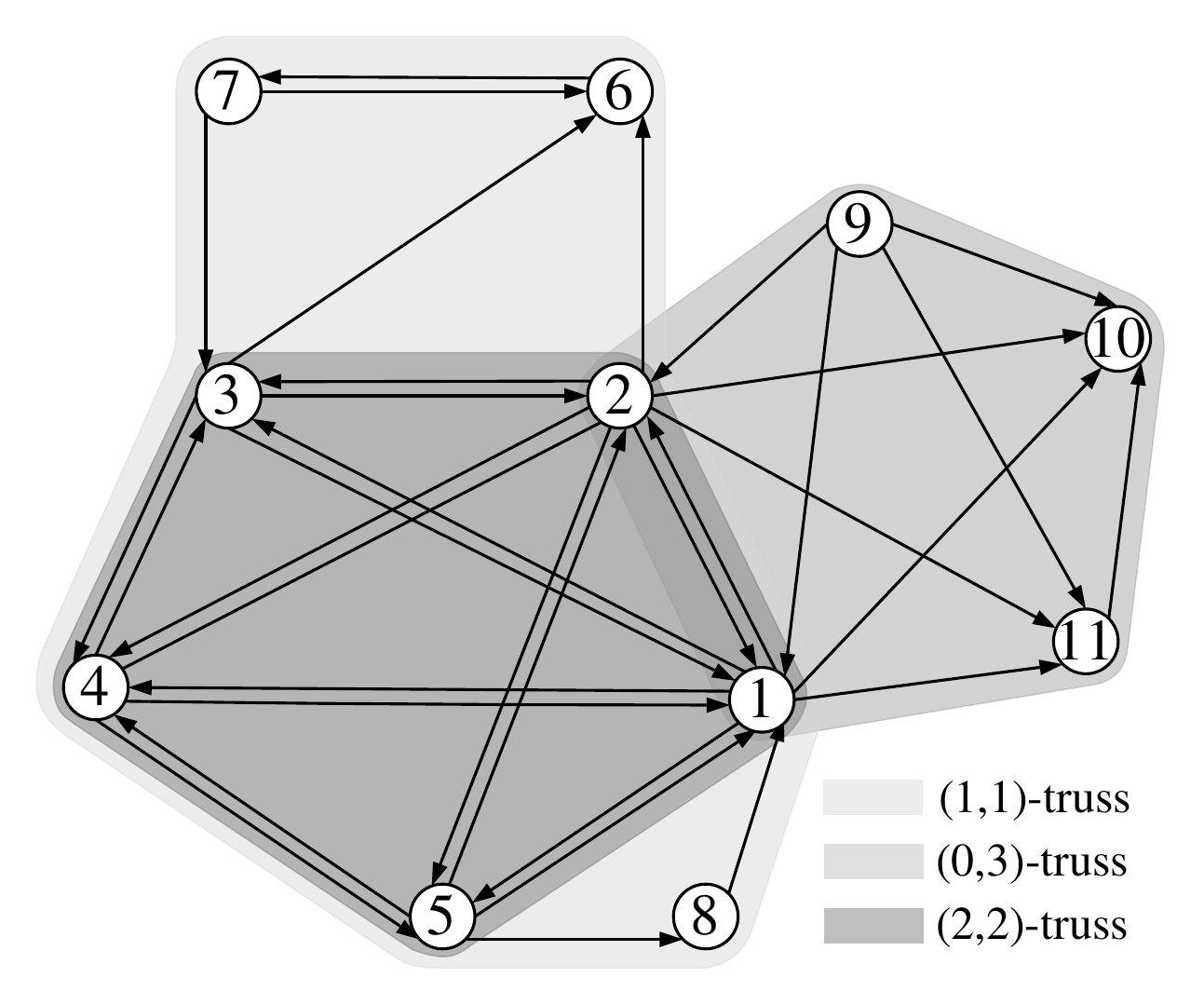}
	\caption{An example of M-D-truss in directed graph.}
	\label{fig_1}
\end{figure}

Thus, this paper introduces an innovative summarized graph\cite{liu2018graph} indexing approach to the community problem in large directed graphs. Specifically, we proposed a novel concept called D-truss-connected, which captures the inherent relationships among edges in D-truss communities. Utilizing this novel notion, any directed graph can be partitioned into connected classes that preserve the D-truss information. Then, we devise and construct a D-truss-connected-based index, ConDTruss, which is space-efficient and cost-effective. Finally, we devise a method based on ConDTruss to uncover the M-D-truss that contains the query vertex efficiently set $Q$. Consider an example in Fig 1. If we want to retrieve the M-(2,2)-truss where node ``$4$'' is located, we only need to find its connected classes and decompress them when outputting the result.

We conduct theoretical and experimental analyses to assess the quality and performance of ConDTruss.

We summarize the key contributions in this paper:

1. We introduce a new concept called D-truss-connected, which captures the inherent relationships among edges in D-truss communities. Utilizing this novel notion, we can divide any directed graph into connected classes for efficient M-D-truss finding.

2. We devise and construct a space-efficient and cost-effective index, ConDTruss. The M-D-truss find can be conducted directly on ConDTruss without visits to the original graph, which is theoretically vintage.

3. We conduct comprehensive experimental studies on various large directed graphs. Experimental results verify the proposed algorithm's effectiveness.

Here, we introduce the organizational structure of this paper. Section 2 provides the preliminaries. Section 3 proposes a novel concept called D-truss-connected and the ConDTruss. The results of the experiments are showcased in Section 4, while Section 5 gives the conclusion of this paper.

Given a directed graph $G_d$ = ($V_{G_d}$, $E_{G_d}$), $V_{G_d}$ and $E_{G_d}$ are the node set and edge set of $G_d$, respectively. We call it a directed graph. For each edge $e$ $\langle u$, $v\rangle$ $\in$ $E_{G_d}$, $u$ is an in-neighbor of $v$ and $v$ is an out-neighbor of $u$. For each node $v$ in $G_d$, the in-degree(out-degree) of $v$ is the number of the in-neighbor (out-neighbor) of $v$, denoted as $\mathrm{degree}_{G_d}^+$($v$) ($\mathrm{degree}_{G_d}^-$($v$)). The degree of $v$ is $|\mathrm{degree}_{G_d}^+$($v$)$|$ + $|\mathrm{degree}_{G_d}^-$($v$)$|$, i.e., $|\mathrm{degree}_{G_d}$($v$)$|$.

{\bf{Definition 1 (Cycle-Support).}} The cycle-support of an edge $e$ = $\langle u$, $v\rangle$ in $E_{G_d}$ $\in$ $G_d$ is defined as $|$\{$w$ in $V_{G_d}$ : $\triangle^\mathrm{C}_{uvw}$ in $G_d$\}$|$, i.e,  $\mathrm{csup}_{G_d}(e)$.

{\bf{Definition 2 (Flow-Support).}} The flow-support of an edge $e$ = $\langle u$, $v\rangle$ in $E_{G_d}$ $\in$ $G_d$ is defined as $|$\{$w$ in $V_{G_d}$ : $\triangle^\mathrm{F}_{uvw}$ in $G_d$\}$|$, i.e,  $\mathrm{fsup}_{G_d}(e)$.

{\bf{Definition 3 (D-truss).}} For a subgraph $H_s$ in $G_d$, $H_s$ is a D-truss($k_c$, $k_f$), if $\forall e$ $\in$ $E_{H_s}$, $\mathrm{csup}_{H_s}(e)$ $\ge$ $k_c$ and $\mathrm{fsup}_{H_s}(e)$ $\ge$ $k_f$.

\section{PRELIMINARIES}
\begin{table}
	\caption{Symbols and interpretations used in this paper\label{tab:table1}}
	\centering
	\begin{tabular}{|c|m{4cm}<{\centering}|c|}
		\hline
		Notation & Description\\
		\hline
		$G_d$ = ($V_{G_d}$, $E_{G_d}$) & A directed, simple graph $G_d$\\
		\hline
		$G_{s}$ = ($V_{S}$, $E_{S}$) & The summarized graph of $G_d$\\
		\hline
		$\mathrm{N}_{G_d}(v)$ & The neighbors of $v$ $\in$ $V_{G_d}$\\
		\hline
		$\mathrm{deg}_{G_d}(v)$ & The degree of $v$ $\in$ $V_{G_d}$\\
		\hline
		$\mathrm{csup}_{G_d}(e)$ & The circle-support of $e$ $\in$ $E_{G_d}$\\
		\hline
		$\mathrm{fsup}_{G_d}(e)$ & The flow-support of $e$ $\in$ $E_{G_d}$\\
		\hline
		$\triangle_{wvu}$ & A triangle constituted by vertices $w, v, u$\\
		\hline
		$\triangle_{wvu}^\mathrm{C}$ & A circle-triangle constituted by vertices $w, v, u$\\
		\hline
	\end{tabular}
\end{table}

\section{Method}

This paper proposes a novel index to find the M-D-truss to systematically overcome the limitations of D-truss methods and achieve a more efficient CS. First, we introduce a D-truss-connected relation to capture edges with similar characteristics in social networks; then, we build a concise and efficient index, ConDTruss, based on this relation to find the M-D-truss.

\subsection{D-truss-connected}

Before starting our work, We need a pre-step to calculate the skyline trussesses of all edges in the input graph.

For an edge $e$ $\in$ $E_{G_d}$, the edge trussness of $e$ is defined as Definition 1.

{\bf{Definition 4 (Edge Trussness).}} Given an edge $e$ $\in$ $E_{G_d}$, ($k_c$, $k_f$) is a trussness of $e$ while a ($k_c$, $k_f$)-truss contain $e$, i.e., T($e$) = {($k_c$, $k_f$)}.

It is important to know that an edge can be contained in multiple D-trusses, resulting in multiple trussnesses for that edge.

{\bf{Definition 5 (Trussness Dominance).}} Given two trussnesses ($k_c^1$, $k_f^1$) and ($k_c^2$, $k_f^2$) of an edge $e$, trussness ($k_c^1$, $k_f^1$) dominates trussness ($k_c^2$, $k_f^2$), denoted as ($k_c^2$, $k_f^2$) $\prec$ ($k_c^1$, $k_f^1$), if: $k_c^1$ $>$ $k_c^2$ and $k_f^1$ $\ge$ $k_f^2$; or (2) $k_c^1$ $\ge$ $k_c^2$ and $k_f^1$ $>$ $k_f^2$.

Note that, if there are two trussnesses ($k_c^1$, $k_f^1$) and ($k_c^2$, $k_f^2$) of $e$, $k_c^1$ $\ge$ $k_c^2$ and $k_f^1$ $\ge$ $k_f^2$, we donated it as ($k_c^2$, $k_f^2$) $\preceq$ ($k_c^1$, $k_f^1$).

{\bf{Definition 6 (Skyline Trussness).}} For an edge $e$ and its trussnesses T($e$) = {($k_c^1$, $k_f^1$), ($k_c^2$, $k_f^2$), ... , ($k_c^n$, $k_f^n$)}, the skyline trussness of $e$ is the trussnesses that are not dominated by others, i.e., ST($e$). Formally, ST($e$) = {($k_c^i$, $k_f^i$) $\in$ $T(e)$: not exists ($k_c^j$, $k_f^j$) $\in$ T($e$), s.t., ($k_c^i$, $k_f^i$) $\prec$ ($k_c^j$, $k_f^j$)}.

We can utilize D-truss decomposition\cite{liu2020truss}, to calculate the skyline trussnesses of every edge in $G_d$.  The outputs of D-truss decomposition are the skyline trussnesses of each edge in $E_{G_d}$. The time complexity(TC) of Algorithm 1 is $O$(min\{$k_{cmax}$, $k_{fmax}$\} $\cdot$ $|E_G|^{1.5}$) and its space complexity(SC) is $O$(min\{$k_{cmax}$, $k_{fmax}$\} $\cdot$ $|E_G|$).

{\bf{Example 2.}} We utilize D-truss decomposition to calculate skyline trussnesses for all edges in the directed graph $G_d$. The results are depicted in Fig 2(a), where edges with distinct skyline trussnesses are visualized in varying colors.

\begin{figure*}[!t]
	\centering
	\subfloat[Edge-trussness]{
		\centering
		\includegraphics[width=2.5in]{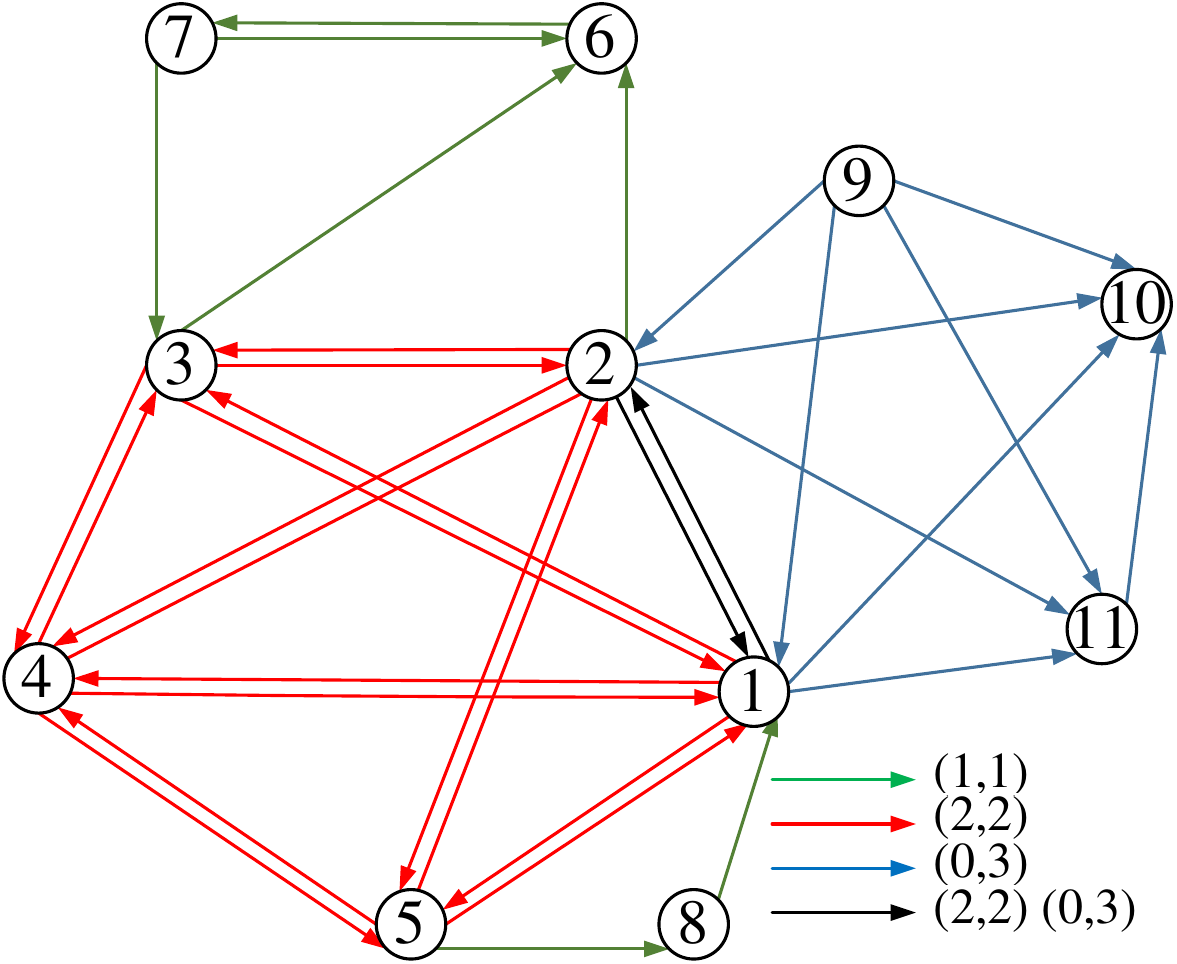}
	}%
	\subfloat[Construct supernodes]{
		\centering
		\includegraphics[width=2.5in]{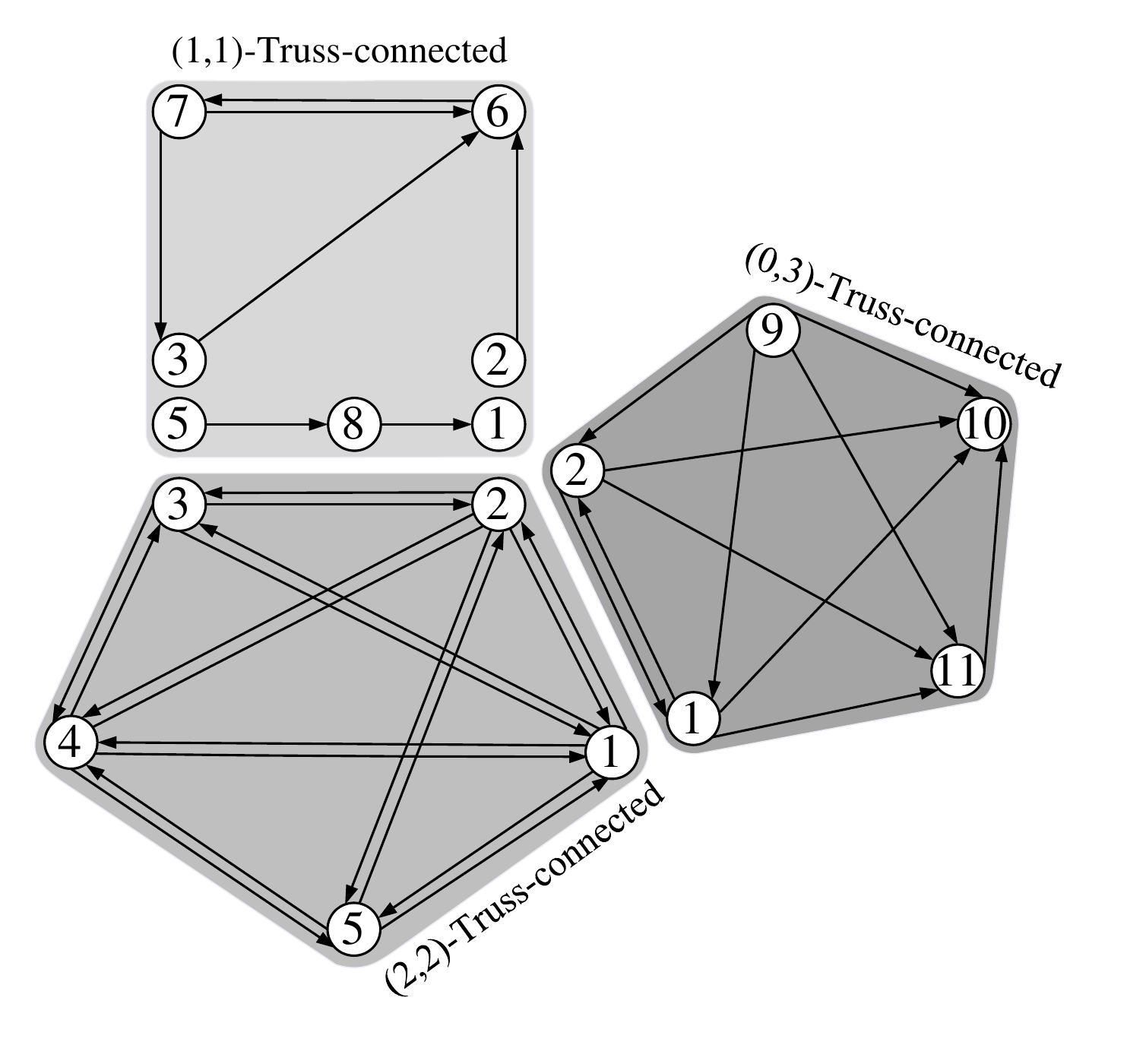}
	}%
	\centering
	\subfloat[Summarized graph]{
		\centering
		\includegraphics[width=2in]{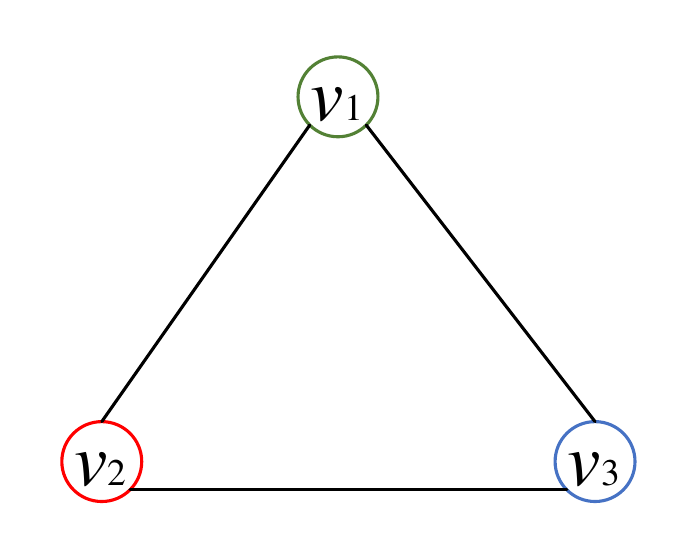}
	}%
	\centering
	\caption{An example of construct summarized graph. (a)Calculate skyline trussnesses for each edge $e\in E_{G_d}$ and partition directed graph $G_d$ into connected classes which preserve D-truss information. (b)Each connected class is represented by a s-node. (c)Construct the summarized graph of $G_d$. And each s-edge depicts the connections between supernodes.}
\end{figure*}

From Fig 2(a), we can notice that for any two edges $e_1$, $e_2$ in $G_d$, they belong to the M-($k_c$, $k_f$)-truss when they satisfy the following two requirements simultaneously: (1) $\tau$($e_1$) $\succeq$ ($k_c$, $k_f$) and $\tau$($e_2$) $\succeq$ ($k_c$, $k_f$); (2) $e_1$ and $e_2$ can be connected through a series of edges where skyline trussness $\succeq$ ($k_c$, $k_f$). Moreover, the two edges satisfying requirement 2, we call them are ($k_c$, $k_f$) connected. Based on this finding, we propose the notion of D-truss-connected as follow,

{\bf{Definition 7 (D-truss-connected (Also called ($k_c$, $k_f$)-truss-connected)).}} For two edges $e_1$, $e_2$ $\in$ $E_{G_d}$, they are ($k_c$, $k_f$)-truss-connected, i.e., $e_{1}$ $\stackrel{(k_c, k_f)}{\Longleftrightarrow}$ $e_{2}$, if (1)($k_c$, $k_f$) $\in$ ST($e_1$) $\cap$ ST($e_2$), and (2)$e_{1}$ and $e_{2}$ are ($k_c$, $k_f$) connected.

\begin{breakablealgorithm}
	\caption{Summarized Graph Construction}\label{alg:alg2}
	\begin{algorithmic}[1]
		\renewcommand{\algorithmicrequire}{\textbf{Input:}}{\textbf{Input:}} a directed graph $G_d$ = ($V_{G_d}$, $E_{G_d}$)
		
		\renewcommand{\algorithmicensure}{\textbf{Output:}}{\textbf{Output:}} a summarized graph $G_s$ = ($V_s$, $E_s$) of ${G_d}$
		
		\STATE D-Truss Decomposition($G_d$);
		\STATE $st$ $\leftarrow \emptyset$;
		\STATE {\textbf{for}} each $e$ $\in$ $E_{G_d}$ {\textbf{do}}
		\STATE \hspace{0.5cm}{\textbf{if}} $\exists$ T($e$) $\in$ ST($e$) = $d$ {\textbf{then}}
		\STATE \hspace{1cm}$\phi_d$ $\leftarrow$ $\phi_d$ $\cup$ $e$;
		\STATE \hspace{1cm}st $\leftarrow$ $st$ $\cup$ \{$d$\};
		\STATE {\textbf{for}} each $d$ $\in$ $st$ {\textbf{do}}
		\STATE \hspace{0.5cm}{\textbf{for}} each $e$ $\in$ $\phi_d$ {\textbf{do}}
		\STATE \hspace{1cm}$\phi_d.e$.$L_id$ $\leftarrow$ $\emptyset$;
		\STATE Id $\leftarrow$ 0;
		\STATE {\textbf{for}} each $d$ in $st$ ($d$ is not dominate by others) {\textbf{do}}
		\STATE \hspace{0.5cm}{\textbf{for}} each $e$ $\in$ $E_{G_d}$ {\textbf{do}}
		\STATE \hspace{1cm}$e$.visited $\leftarrow$ False;
		\STATE \hspace{0.5cm}{\textbf{while}} $\exists e$ $\in$ $\phi_d$ {\textbf{do}}
		\STATE \hspace{1cm}$e$.visited $\leftarrow$ True;
		\STATE \hspace{1cm}$L$ $\leftarrow$ $\emptyset$;
		\STATE \hspace{1cm}Create a supernode $\nu$ with $\nu$.Id $\leftarrow$ Id + 1;
		
		\STATE \hspace{1cm}$V_s$ $\leftarrow$ $V_s$ $\cup$ \{$\nu$\};
		\STATE \hspace{1cm}$L$.append($e$);
		\STATE \hspace{1cm}{\textbf{while}} $|L|$ $\neq$ 0 {\textbf{do}}
		
		\STATE \hspace{1.5cm}$e$$\langle u$, $v\rangle$ $\leftarrow$ $L$.pop();
		\STATE \hspace{1.5cm}{\textbf{if}} $\exists$ T($e$) $\in$ ST($e$) = $d$ {\textbf{then}}
		\STATE \hspace{2cm}$\nu$ $\leftarrow$ $\nu$ $\cup$ \{$e$\};
		\STATE \hspace{1.5cm}{\textbf{for}} each id $\in$ $e$.$L_id$ {\textbf{do}}
		\STATE \hspace{2cm}$E_s$ $\leftarrow$ $E_s$ $\cup$\{$\langle \mu$, $\nu \rangle$\};
		\STATE \hspace{1.5cm}{\textbf{for}} each incident edge $e'$ of $e$ {\textbf{do}}
		\STATE \hspace{2cm}ProcessEdge1($e'$); 
		\STATE \hspace{1.5cm}$\phi_d$ $\leftarrow$ $\phi_d$ - $e$; ST($e$) $\leftarrow$ ST($e$) - $d$;
		\STATE \hspace{1.5cm}{\textbf{if}} $|$ST($e$)$|$ = 0 {\textbf{then}}
		\STATE \hspace{2cm}$E_{G_d}$ $\leftarrow$ $E_{G_d}$ - $e$;
		\STATE \hspace{0.5cm}$st$ $\leftarrow$ $st$ - $d$;
		\STATE {\textbf{Return}} $G_s$ = ($V_s$, $E_s$) 
		
		\renewcommand{\algorithmicensure}{\textbf{Procedure ProcessEdge($e$):}}{\textbf{Procedure ProcessEdge($e$):}} 
		\STATE {\textbf{if}} $d$ $\in$ ST($e$) and $\phi_d.e$.visited = False {\textbf{then}}
		\STATE \hspace{0.5cm}$e$.visited = True;
		\STATE \hspace{0.5cm}$L$.append($e$);
		\STATE {\textbf{if}} $\exists$ $\tau$ $\in$ ST($e$) $\npreceq$ $d$ and  Id $\notin$ $\phi_d.e$.$L_id$ {\textbf{then}}
		\STATE \hspace{0.5cm}$e$.visited = True;
		\STATE \hspace{0.5cm}($e$).$L_id$ $\leftarrow$ $\phi_d.e$.$L_id$ $\cup$ \{Id\}
		\STATE \hspace{0.5cm}{\textbf{if}} $\exists$ $\tau$ $\in$ ST($e$) $\succ$ $d$ {\textbf{then}}
		\STATE \hspace{1cm}$L$.append($e$);
	\end{algorithmic}
	\label{alg2}
\end{breakablealgorithm}

\subsection{Index construction}
Afterward, we constructed the ConDTruss to maintain the skyline trussness and adjacency information in $G_d$. We regard each connected class as a supernode and establish superedges between them according to their connected relations. The detailed process is described in Algorithm 1.

For a directed graph $G_d$, we develop the ConDTruss for $G_d$ based on D-truss-connected in Algorithm 1. During the initialization process (Lines 1-10), the algorithm invokes D-truss decomposition to calculate the skyline trussness for each edge $e$ $\in$ $E_{G_d}$(line 1). We reassign the edges according to the skyline trussnesses of each edge to a different set $\Phi_d$ and record all skyline trussnesses to $st$ (lines 2-6). Given an edge $e$ $\in$ $\Phi_d$, we preserve an auxiliary data structure $L_id$, which is a set of supernode labels, where each label corresponds to a supernode that has been previously explored, $\mu$, where T($\mu$) does not dominate $k$(denotes as $d$ $\npreceq$ T($\mu$)), $\mu$ is connected to the current supernode $\nu$, T($\nu$) = $d$. The set $\Phi_d.e$.$L_id$ is initialized as empty (line 9). Given a value $d$ in $st$, if others do not dominate $d$, the algorithm examines edges in $\Phi_d$. (Lines 11-30). When an edge $e$ $\in$ $\Phi_d$ is selected, a new supernode $\nu$ will be set to represent the connected class of $e$ (Lines 17-18). We identify all edges that are D-truss D-Truss-connected to $e$ and add them to the supernode $\nu$ by BFS (lines 20-27). During the exploration process, we also examine if there exists a supernode $\mu$ in $\phi_d.e$.$L_id$ that satisfies T($\nu$) $\npreceq$ T($\mu$) and $\mu$ is connected to $\nu$ through edge $e$. If such a supernode $\mu$ is found, we set a superedge ($\mu$, $\nu$) in the index (Lines 24-25). In the end, we will get a summarized graph index made of $V_s$ and $E_s$.

{\textbf{Complexity Analysis.}} In Algorithm 1. In the initialization process (Lines 1-11), the D-truss decomposition takes $O$(min\{$k_{cmax}$, $k_{fmax}$\} $\cdot$ $E_{G_d}^{1.5}$) time. In the summarized graph development process (Lines 12-36), given an edge $e$ $\in$ $E_{G_d}$, and $e$ is in $n$ sets of $\phi_d$ ($n$ = $|$ST($e$)$|$). For each edge $e$ $\in$ $\phi_d$, we identify all edges D-Truss-connected to $e$ by examining all incident edges of $e$, then $e$ is removed from $\phi_d$. Therefore, each $e$ $\in$ $E_{G_d}$ is examined $n$ times. The procedure ProcessEdge1 and ProcessEdge2 takes $O$(1)time. Thus, the TC of Algorithm 2 is $O$(min\{$k_{cmax}$, $k_{fmax}$\}$\cdot$ $|E_{G_d}|^{1.5}$ + $n$ $\cdot$ $|E_{G_d}|$). Moreover, each $e$ $\in$ $E_{G_d}$ can be in $n$ supernodes, so the SC of Algorithm 2 is $n$ $\cdot$ $|E_{G_d}|$. 

{\bf{Example 3.}} The ConDTruss of the directed graph $G_d$ is illustrated in Fig 2(c). It consists of 3 supernodes, each representing a D-truss-connected class for the edges in $G_d$. For instance, the supernode $\nu_3$ corresponds to a (0, 3)-truss community comprising 11 edges. These edges are connected and share the same skyline trussness of (0, 3). Additionally, it contains three superedges that depict the connectivity between supernodes.

\subsection{ConDTruss-based maximal D-truss find.}

After the ConDTruss is developed from $G_d$, we can efficiently find the M-D-truss directly on the ConDTruss.
The retrieval process is as described in Algorithm 2.

\begin{breakablealgorithm}
	\caption{ConDTruss-based M-D-truss find.}\label{alg:alg6}
	\begin{algorithmic}[1]
		\renewcommand{\algorithmicrequire}{\textbf{Input:}}{\textbf{Input:}} $G_s$=($V_s$, $E_s$), a query node set $Q$, $k_c$ and $k_f$
		
		\renewcommand{\algorithmicensure}{\textbf{Output:}}{\textbf{Output:}} the M-D-truss containing $q$
		\STATE {\textbf{for}} each $\nu$ $\in$ $V_s$ {\textbf{do}}
		\STATE \hspace{0.5cm}$\nu$.visited $\leftarrow$ False;
		\STATE $D_m$ $\leftarrow$ $\emptyset$;
		\STATE {\textbf{while}}$|L_Q|$ $\neq$ 0 {\textbf{do}}
		\STATE \hspace{0.5cm}$\nu$ $\leftarrow$ $L_Q$.pop;
		\STATE \hspace{0.5cm}$\nu$.visited = True;
		\STATE \hspace{0.5cm}$D_m$ $\leftarrow$ $D_m$ $\cup$\{$e|e$ $\in$ $\nu$\};
		\STATE \hspace{0.5cm}{\textbf{for}} each $\mu$ $\in$ N($\nu$) {\textbf{do}}
		\STATE \hspace{1cm}{\textbf{if}} $\tau$($\mu$) $\succeq$ ($k_c$, $k_f$) and $\mu$.visited = False {\textbf{then}}
		
		\STATE \hspace{1.5cm}$\mu$.visited $\leftarrow$ True;
		\STATE \hspace{1.5cm}$L_Q$.append($\mu$);
		\STATE {\textbf{Return}} $D_m$
	\end{algorithmic}
	\label{alg6}
\end{breakablealgorithm}

Given the ConDTruss of $G_d$, a query node set $Q$, and two integers $k_c$ and $k_f$. Initially, we put each supernode $\nu$ which the query node $q$ in $Q$ belongs to and $\tau$($\nu$) $\succeq$ ($k_c$, $k_f$) into a list $L_Q$. Next, for each supernode $\nu$ $\in$ $L_Q$ with $\tau$($\nu$) $\succeq$ ($k_c$, $k_f$), we traverse $G_s$ in a BFS fashion(lines 4 - 11). For each neighboring supernode $\mu$, if $\mu$ has not been visited and $\tau$($\mu$) $\succeq$ ($k_c$, $k_f$), we add the edges within $\mu$ to the M-D-truss $D_m$ (lines 7). Afterward, we will get a M-D-truss containing $Q$.

{\textbf{Complexity Analysis.}} In Algorithm 2, each edge of $D_m$ is visited only once when decompressed as result. So the TC of Algorithm 3 is $O$($|D_m|$).

{\bf{Example 4.}} Consider the directed graph $G_d$ in Fig 2(a), two integers $k_c$ = 2 and $k_f$ = 2, and a query node $'2'$. We first find the supernode from the summarized graph where $'2'$ is located, which is $\nu_2$. Starting from $\nu_2$, T($\nu_2$) = (2, 2), so the algorithm adds all the edges in $\nu_2$ to the M-D-truss $D_m$. However, $\nu_2$'s neighboring supernodes $\nu_1$ and $\nu_3$ are disqualified because $\tau$($\nu_1$) = (1, 1) and $\tau$($\nu_3$) = (0, 3). Finally, we will get the M-D-truss, $D_m$, as shown in Fig 3.

\begin{figure}[H]
	\centering
	\includegraphics[width=3.5in]{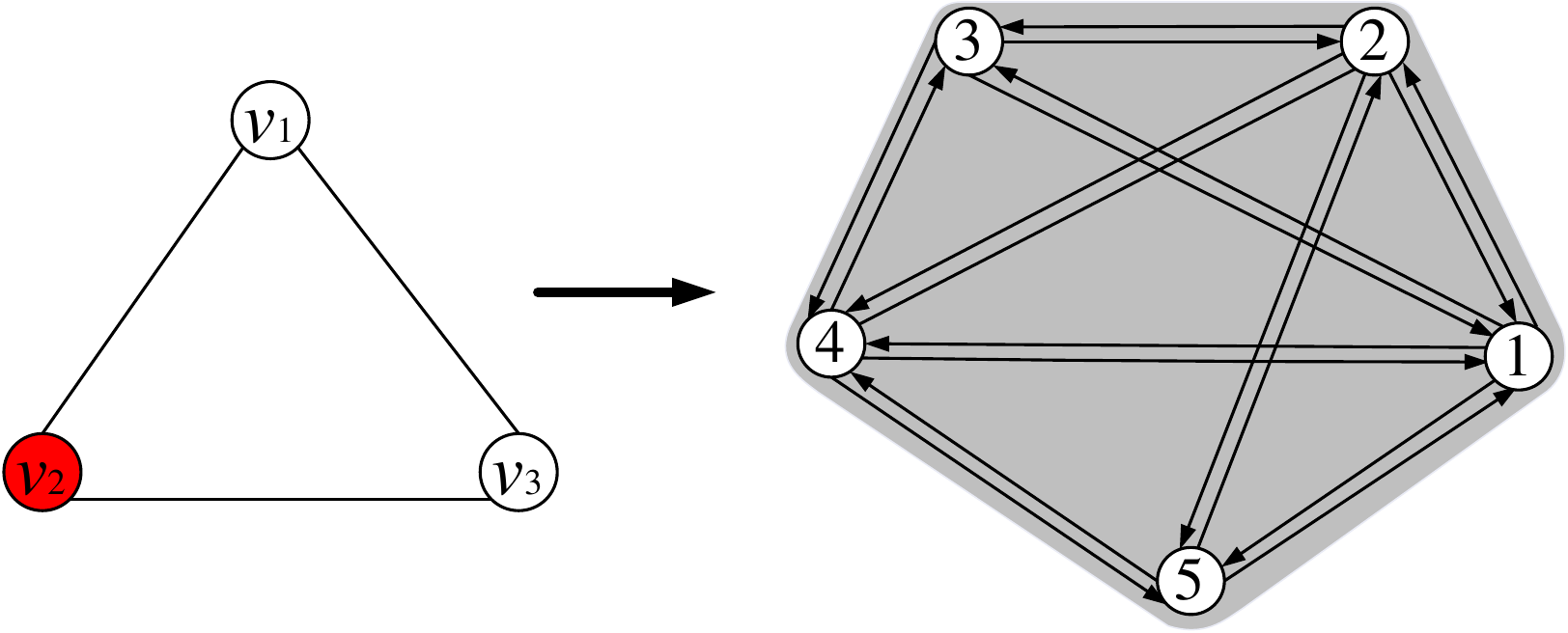}
	\caption{An example of ConDTruss-based M-D-truss find}
	\label{fig_5}
\end{figure}

\section{Experiment}
This section verifies the effectiveness of our proposed index and algorithm. All experiments were conducted on a Windows Server with a six-core CPU running at 2.50 GHz and 32GB of RAM. The algorithms were coded using Python.

Four real-world datasets consisting of directed networks are utilized in our experiments. The statistical information for these networks is summarized in Table 2., including four datasets: EAT, Slashdot, Twitter, and Pokec. EAT is a word association network, and the other three data sets are social relationship networks.

\begin{table}[H]
	\caption{Dataset statistics ($K$ = $10^3$, $M$ = $10^6$)\label{tab:table2}}
	\centering
	\begin{tabular}{|cccccc|}
		\hline
		\multicolumn{1}{|c|}{Datasets} & \multicolumn{1}{l|}{$|V_G|$} & \multicolumn{1}{l|}{$|E_G|$} & \multicolumn{1}{l|}{$d_{max}$} & \multicolumn{1}{l|}{$k_{cmax}$} & \multicolumn{1}{l|}{$k_{fmax}$} \\ \hline

		\multicolumn{1}{|c|}{ETA} & \multicolumn{1}{l|}{23.1K} & \multicolumn{1}{l|}{685K} & \multicolumn{1}{l|}{1,106} & \multicolumn{1}{l|}{3} & \multicolumn{1}{l|}{8}  \\ \hline
		\multicolumn{1}{|c|}{Slashdot} & \multicolumn{1}{l|}{77.4K} & \multicolumn{1}{l|}{905.5K} & \multicolumn{1}{l|}{5,048} & \multicolumn{1}{l|}{33} & \multicolumn{1}{l|}{33}  \\ \hline
		\multicolumn{1}{|c|}{Twitter} & \multicolumn{1}{l|}{81.3k} & \multicolumn{1}{l|}{1.8M} & \multicolumn{1}{l|}{3,758} & \multicolumn{1}{l|}{161} & \multicolumn{1}{l|}{199} \\ \hline
		\multicolumn{1}{|c|}{Pokec} & \multicolumn{1}{l|}{1.6M} & \multicolumn{1}{l|}{30.6M} & \multicolumn{1}{l|}{20,518} & \multicolumn{1}{l|}{18} & \multicolumn{1}{l|}{27} \\ \hline
	\end{tabular}
\end{table}

\subsection{Index construction}

We initiate our experiments by constructing the indexes, which are done offline prior to CS. After constructing indexes, they are stored in the main memory, enabling efficient CS in large graphs. Our experimental analyses emphasize three evaluation metrics:

(1)The time required for index construction.;

(2)The memory size of the index;

(3)The edge compression ratio(ECR): $|E_s|$ / $|E_{G_d}|$.

The experimental results are presented in Table 3.

\begin{table}[H]
	\caption{The time and space required for the DEBI build and the size of the original graph.\label{tab:table3}}
	\centering
	\begin{tabular}{|l|l|l|l|l|}
		\hline
		\multicolumn{1}{|c|}{Graph} & Size(MB) & Index size(MB) & Time(s) & ECR    \\ \hline
		ETA                         & 6.92     & 4.12            & 385.72                  & 0.0009 \\ \hline
		Slashdot                    & 10.49    & 6.85            & 892.36                 & 0.0017 \\ \hline
		Twitter                     & 43.5     & 27.31            & 2368.61                 & 0.0113 \\ \hline
		Pokec                       & 404.3    & 229.76           & 32574.83                & 0.0016 \\ \hline
	\end{tabular}
\end{table}

Table 3 shows that ConDTruss can be efficiently constructed from the original graph, and its size is always smaller than the original graph because each edge in the original graph is compressed in the corresponding supernode. The ECR describes the degree to which the original graph is compressed. The lower the index, the higher the query efficiency of the maximum D-truss.

\subsection{Case study}

Since our method has the same effectiveness as \cite{liu2020truss}, no experiments are performed on the community quality measure. We will perform the case analyses on the EAT in this part of the experiment. We run two queries on node ``$DRINK$'' separately; $ k_c$, $k_f$ values are (0,7) and (0,8) respectively. The results shown in Fig 4 indicate that we can obtain communities with different degrees of tightness by adjusting the $k_c$ or $k_f$. This capability is crucial for personalized CS in analyzing and studying large-scale graphs.

\begin{figure}[H]
	\centering
	\subfloat[(0,7)-truss]{
		\centering
		\includegraphics[width=1.7in]{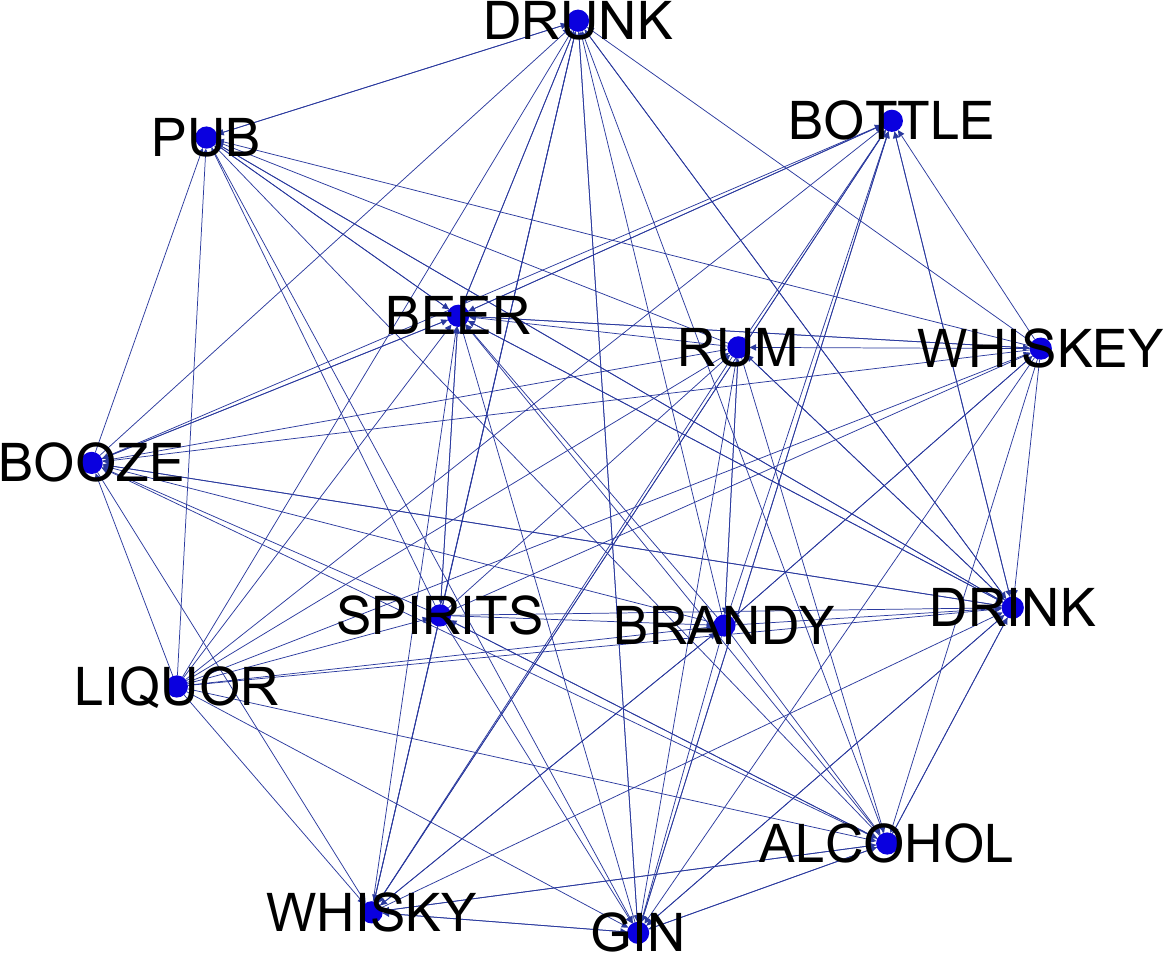}
	}%
	\subfloat[(0,8)-truss]{
		\centering
		\includegraphics[width=1.7in]{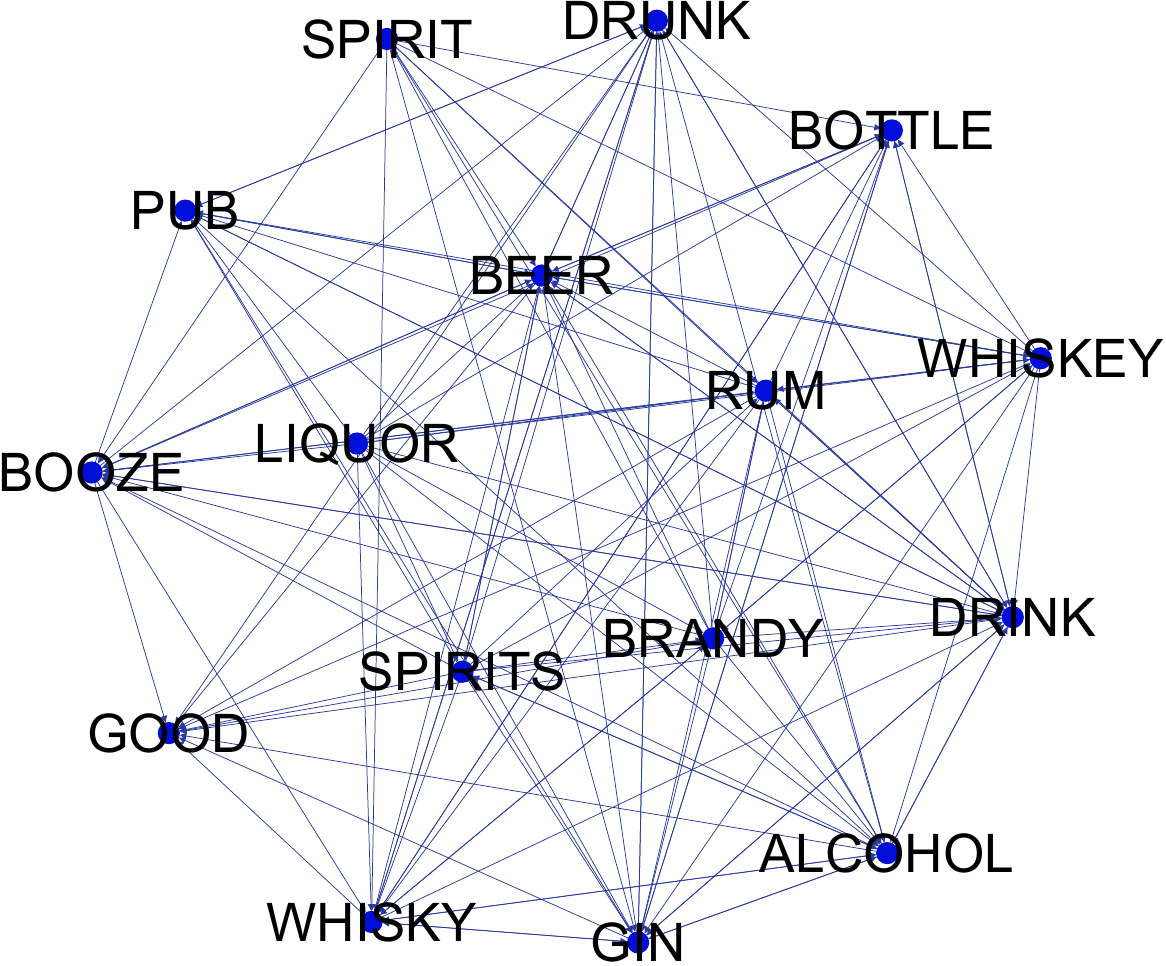}
	}%
	\centering
	\caption{Case study on EAT. (a) A (0, 7)-truss community of ``$DRINK$'' in EAT. (b) A (0, 8)-truss community of ``$DRINK$'' in EAT. }
\end{figure}

\subsection{Performance Evaluation}

\begin{figure*}
	\centering
	\subfloat[EAT]{
		\centering
		\includegraphics[width=1.55in]{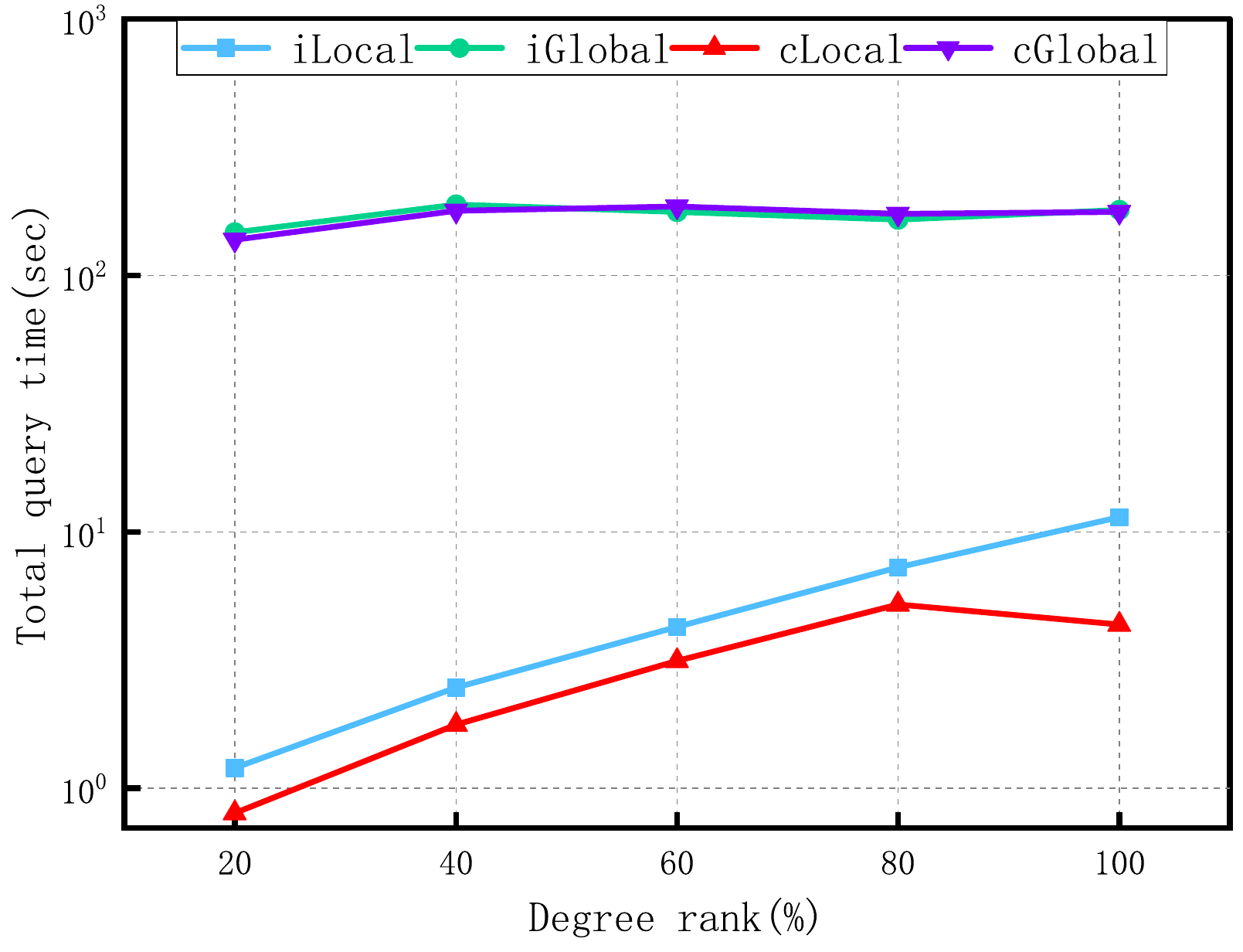}
	}%
	\subfloat[Slash]{
		\centering
		\includegraphics[width=1.55in]{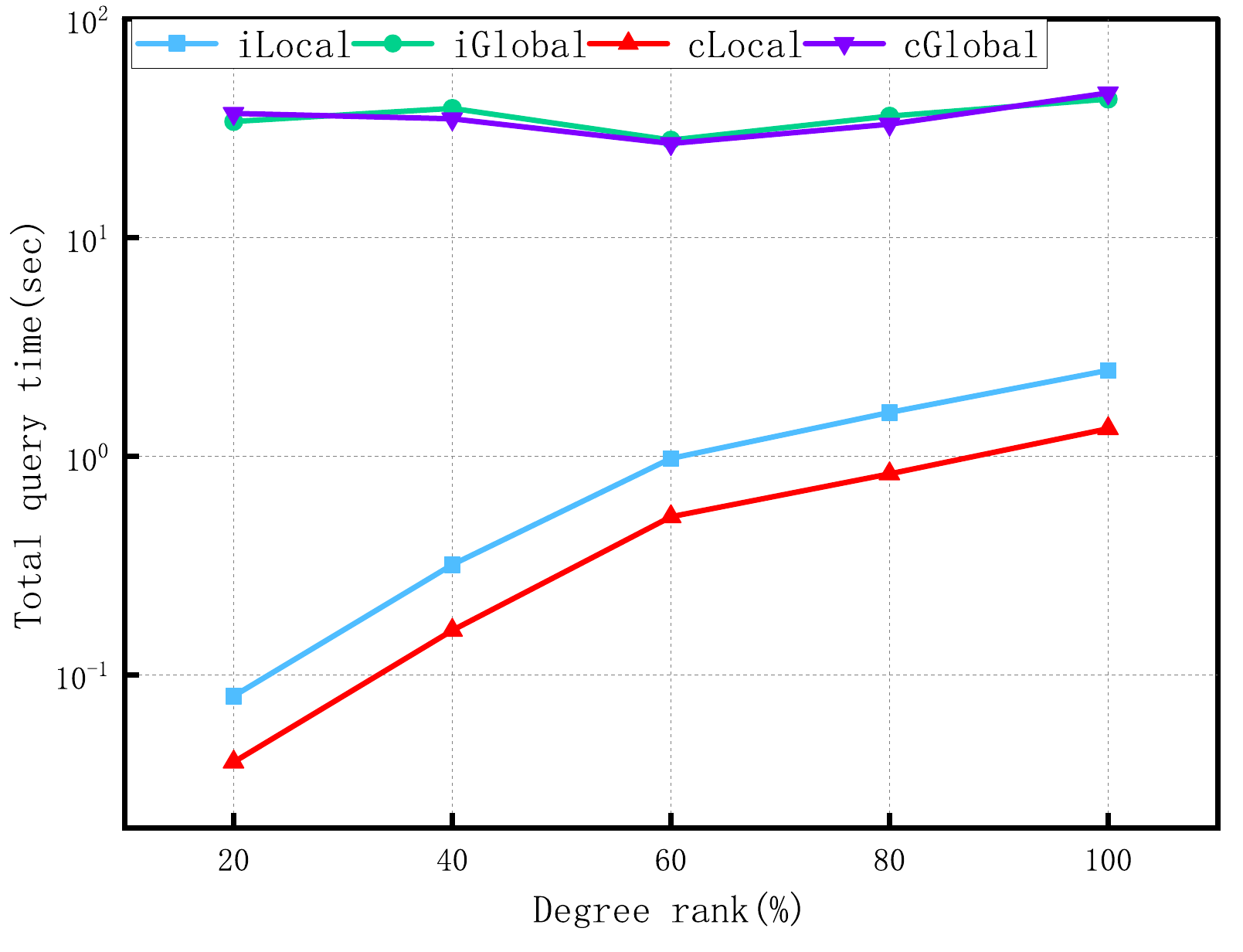}
	}%
	\subfloat[Twitter]{
		\centering
		\includegraphics[width=1.55in]{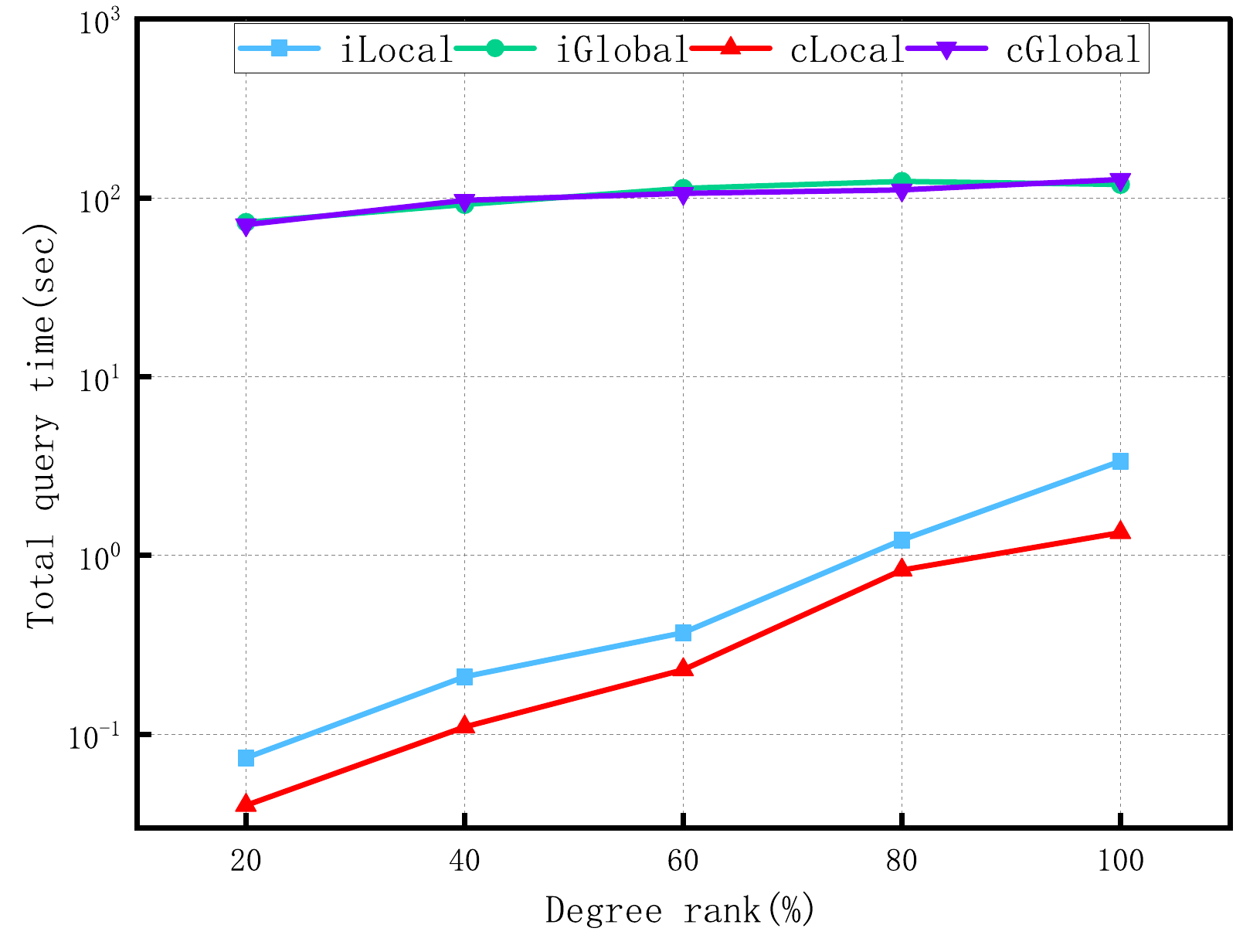}
	}%
	\subfloat[Pokec]{
		\centering
		\includegraphics[width=1.55in]{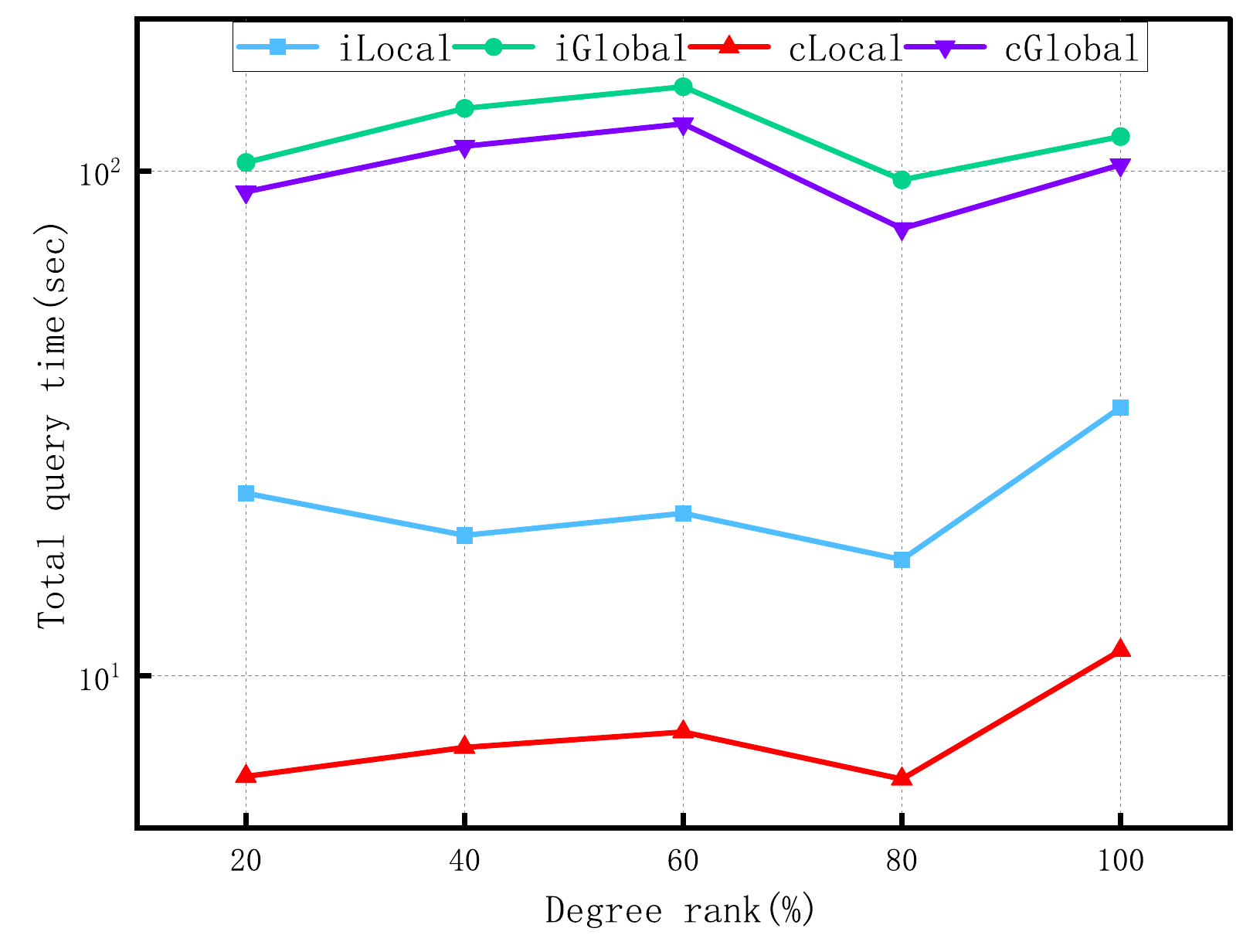}
	}%
	\centering
	\caption{CS performance in different node-degree}
\end{figure*}

\begin{figure*}
	\centering
	\subfloat[EAT]{
		\centering
		\includegraphics[width=1.55in]{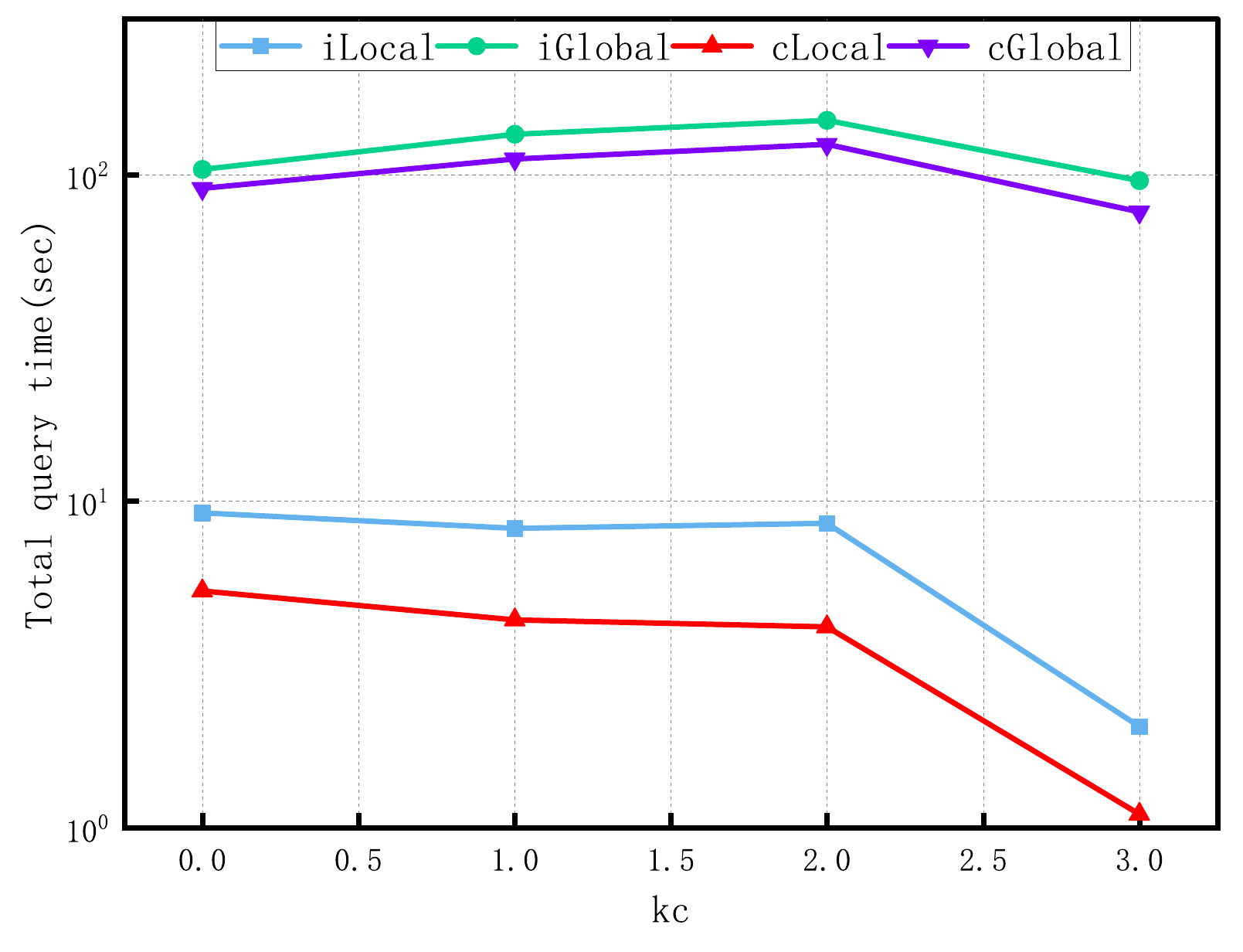}
	}%
	\subfloat[Slash]{
		\centering
		\includegraphics[width=1.55in]{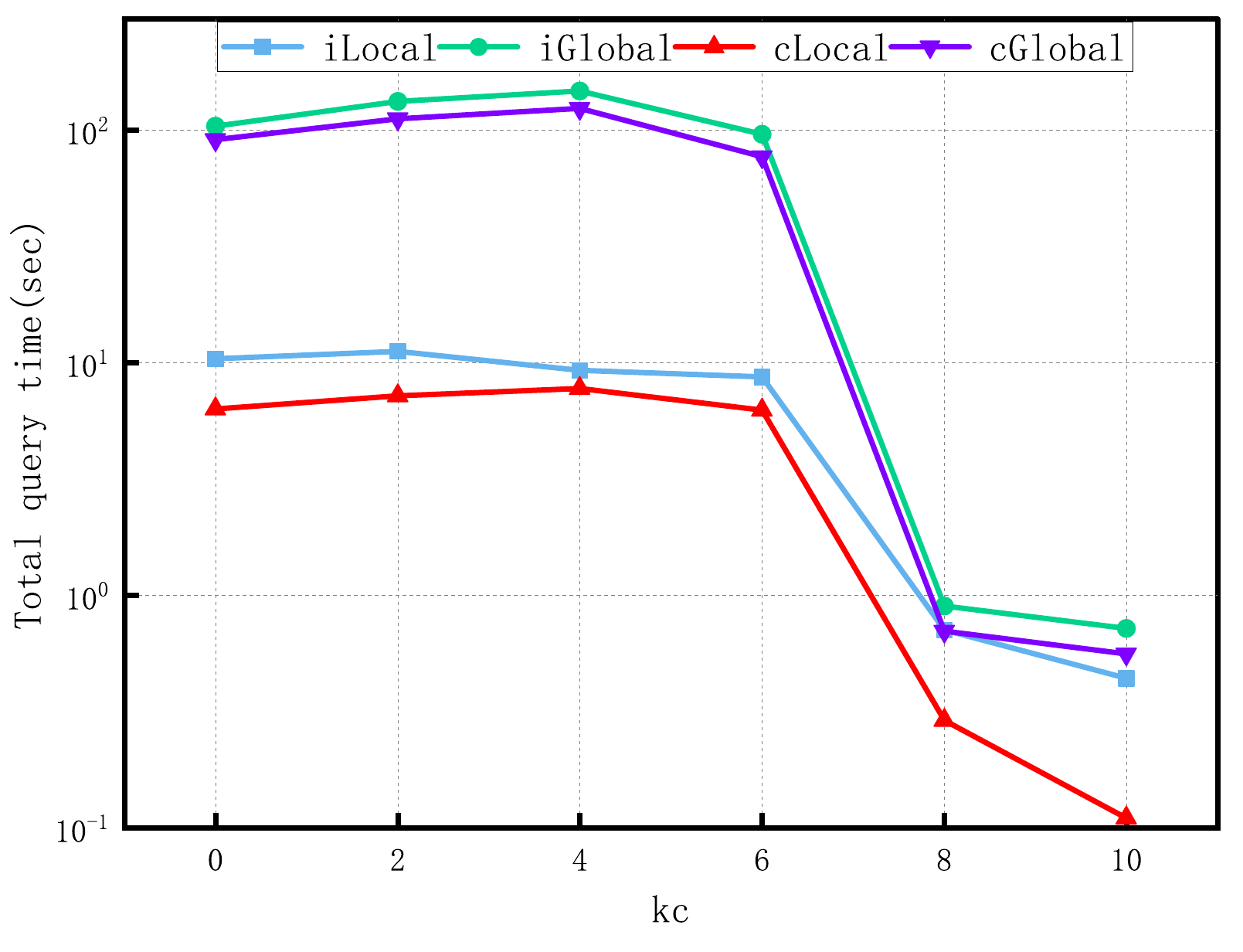}
	}%
	\subfloat[Twitter]{
		\centering
		\includegraphics[width=1.55in]{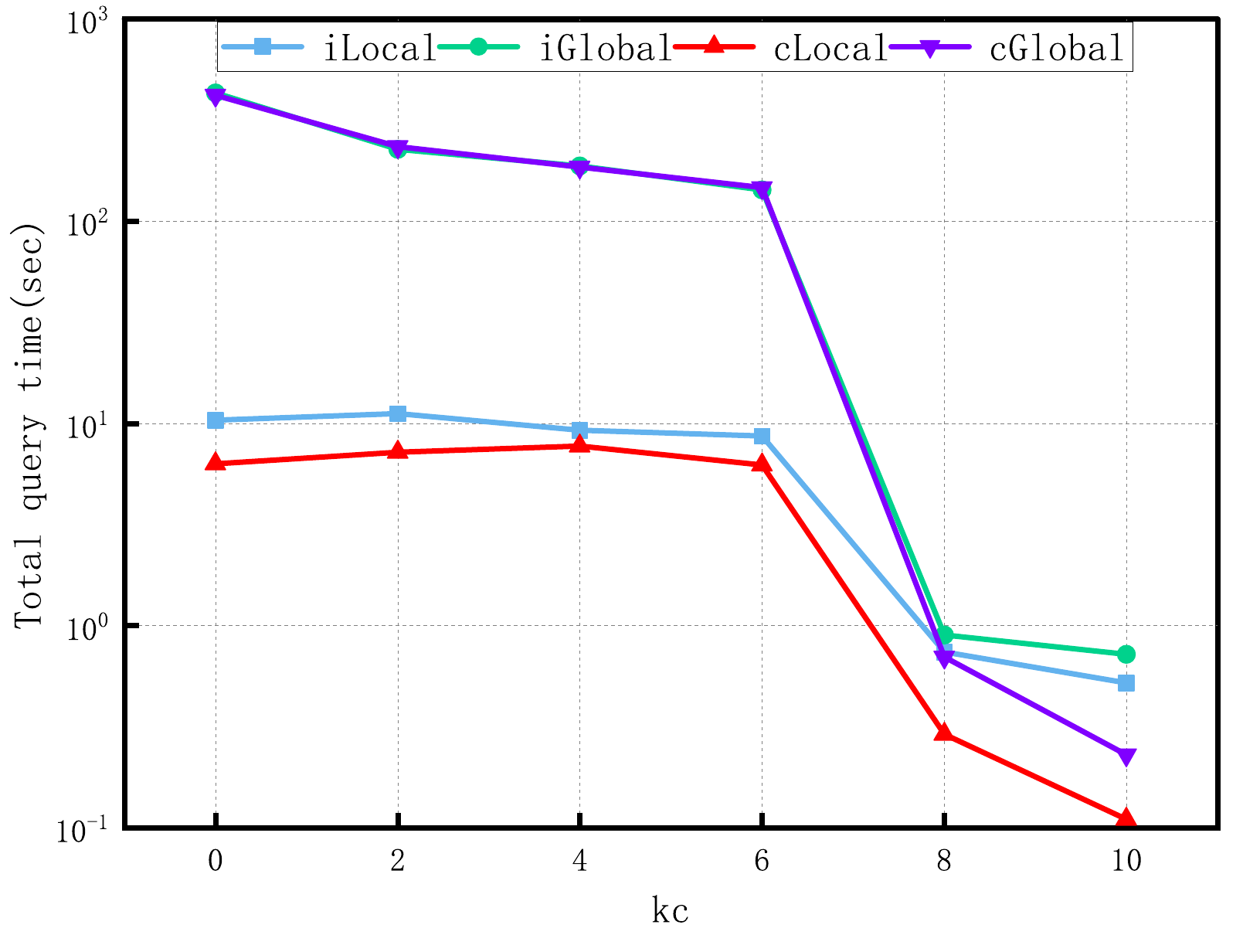}
	}%
	\subfloat[Pokec]{
		\centering
		\includegraphics[width=1.55in]{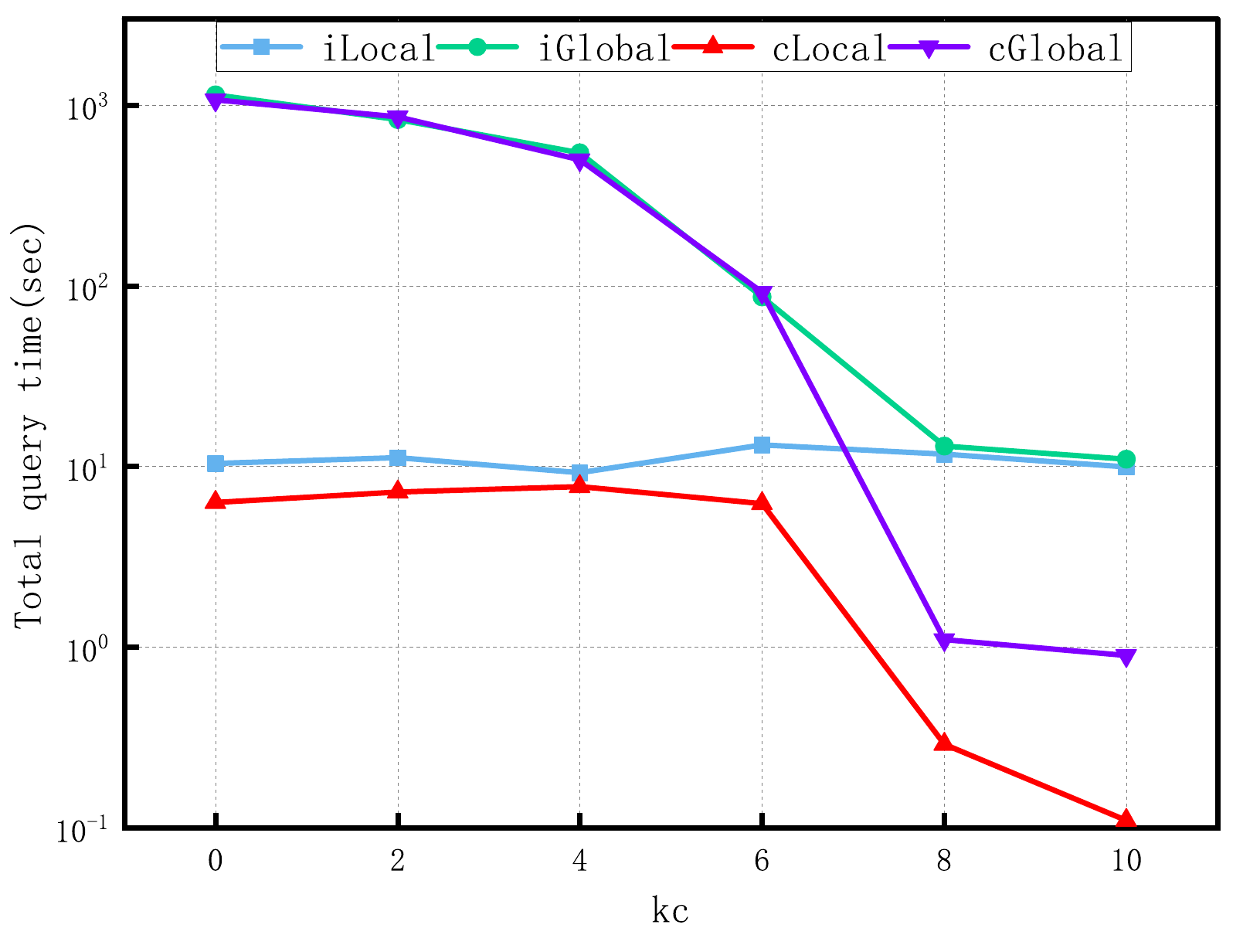}
	}%
	\centering
	\caption{CS performance for different values of $k_c$}
\end{figure*}
\begin{figure*}
	\centering
	\subfloat[EAT]{
		\centering
		\includegraphics[width=1.55in]{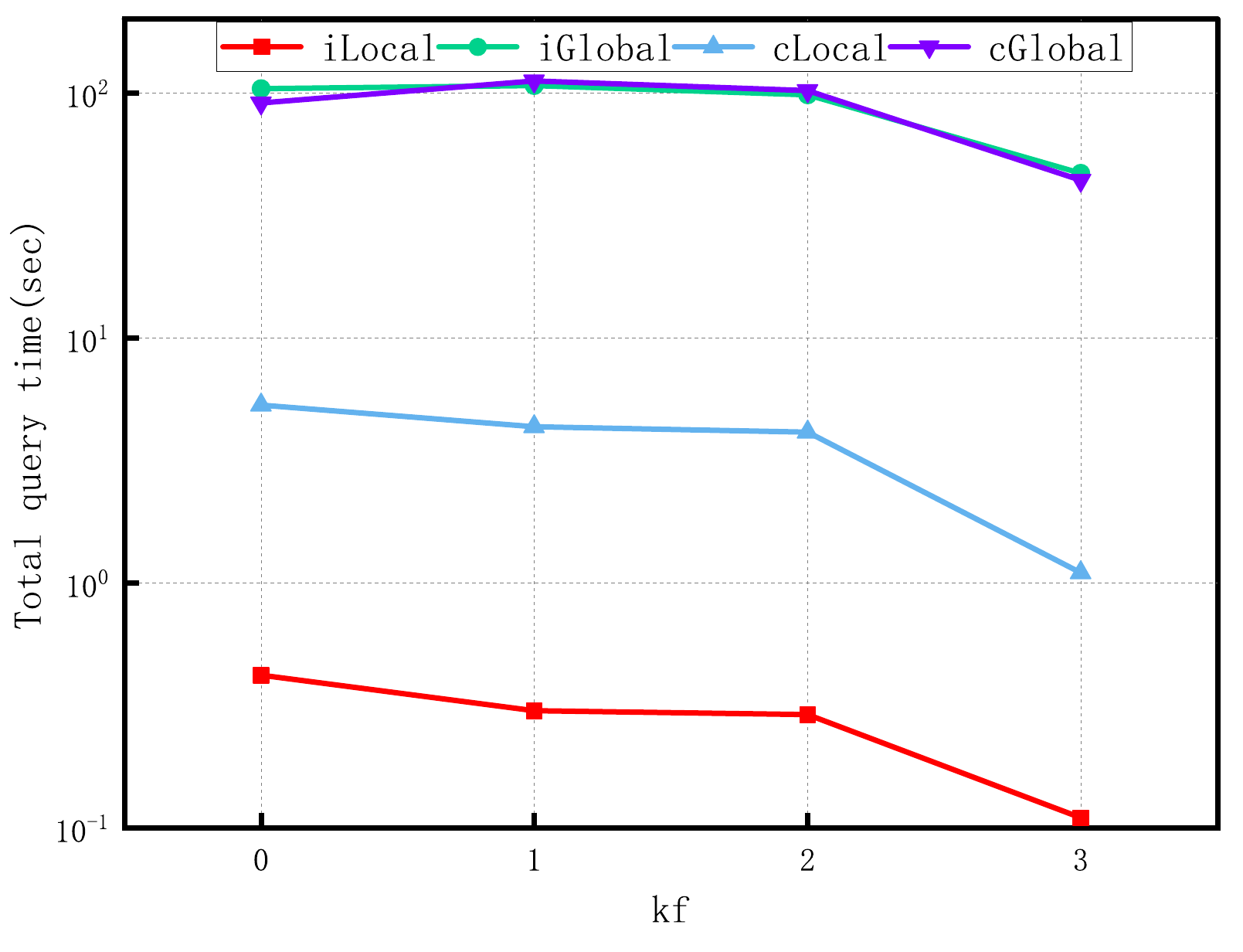}
	}%
	\subfloat[Slash]{
		\centering
		\includegraphics[width=1.55in]{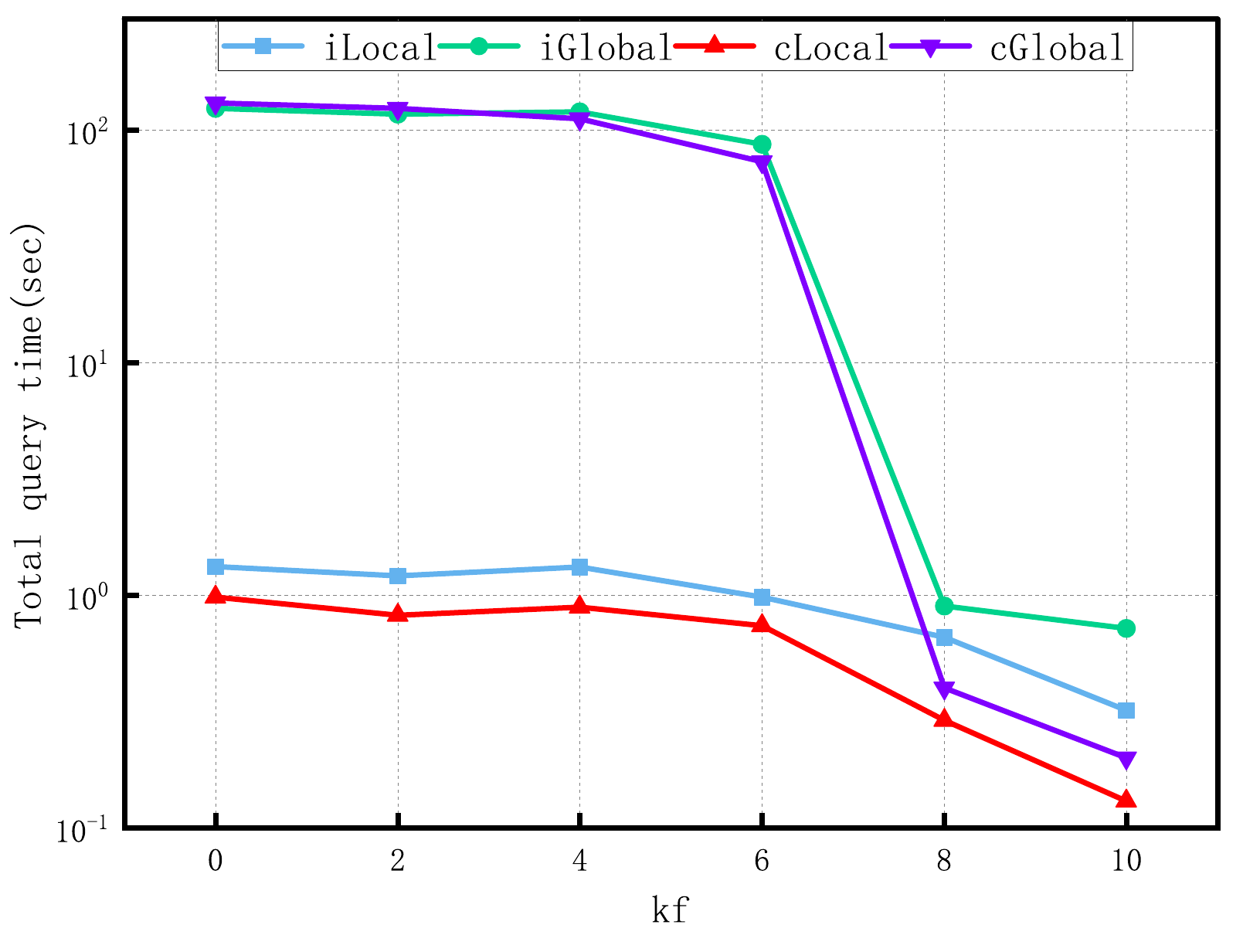}
	}%
	\subfloat[Twitter]{
		\centering
		\includegraphics[width=1.55in]{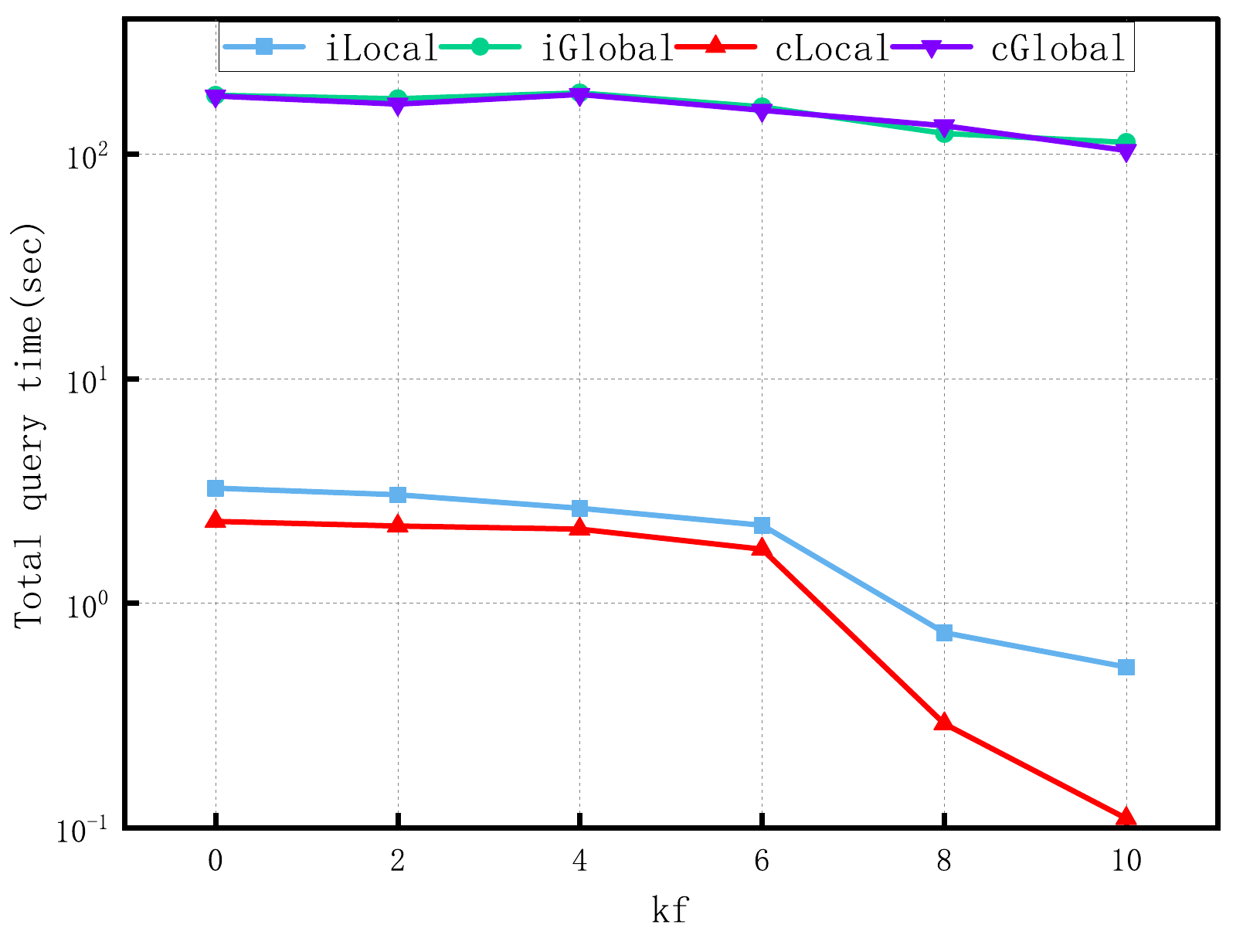}
	}%
	\subfloat[Pokec]{
		\centering
		\includegraphics[width=1.55in]{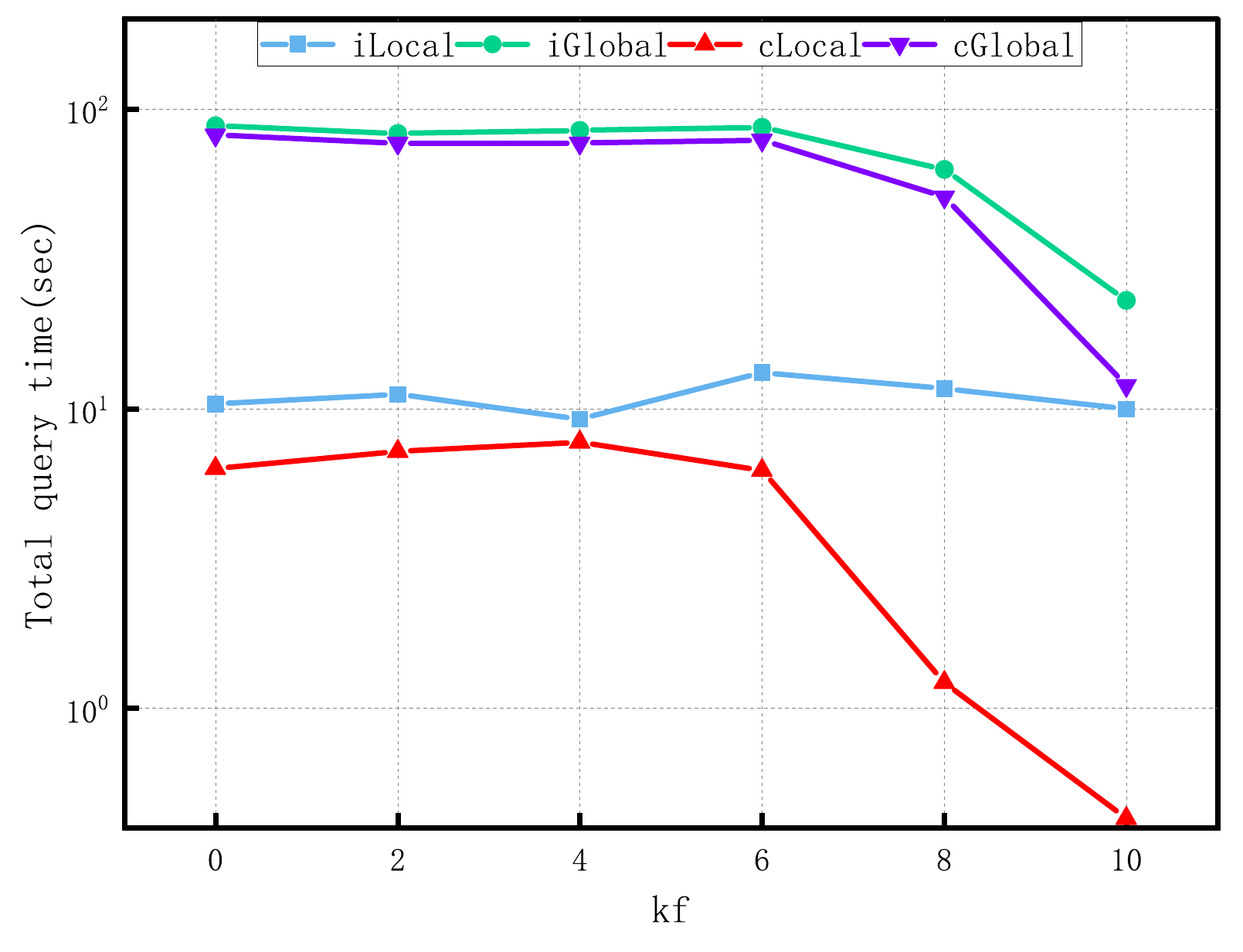}
	}%
	\centering
	\caption{CS performance for different values of $k_f$}
\end{figure*}
\begin{figure*}
	\centering
	\subfloat[EAT]{
		\centering
		\includegraphics[width=1.55in]{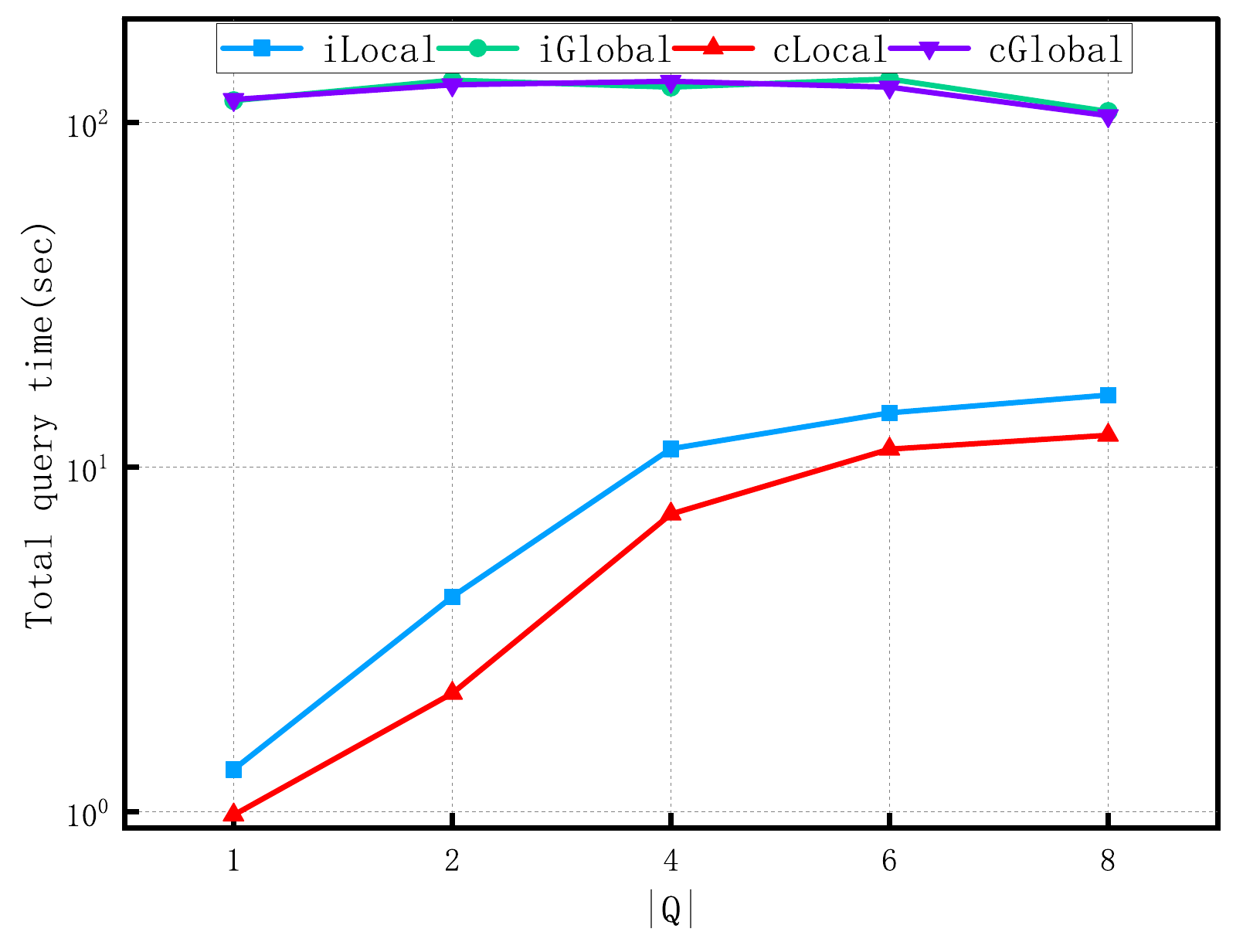}
	}%
	\subfloat[Slash]{
		\centering
		\includegraphics[width=1.55in]{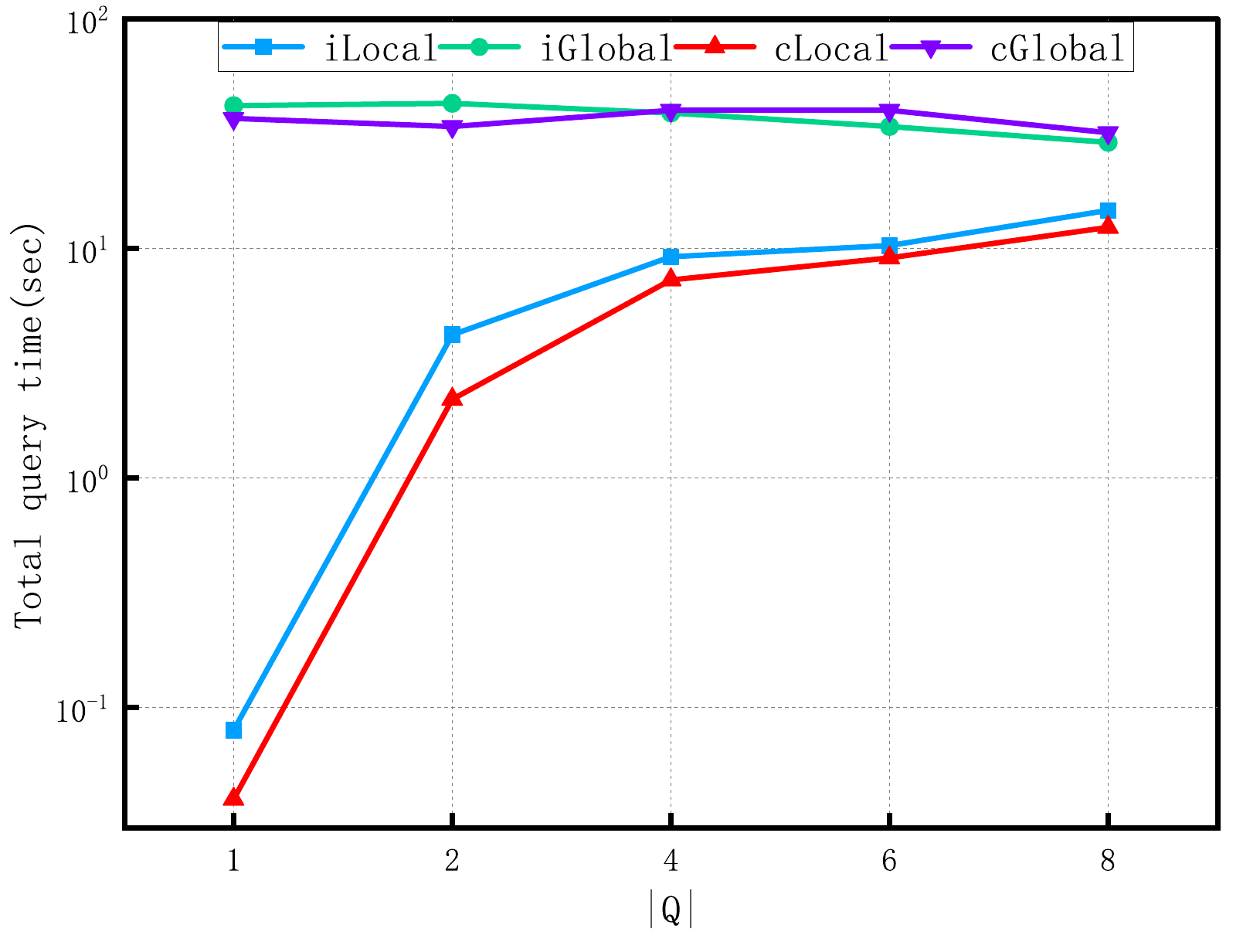}
	}%
	\subfloat[Twitter]{
		\centering
		\includegraphics[width=1.55in]{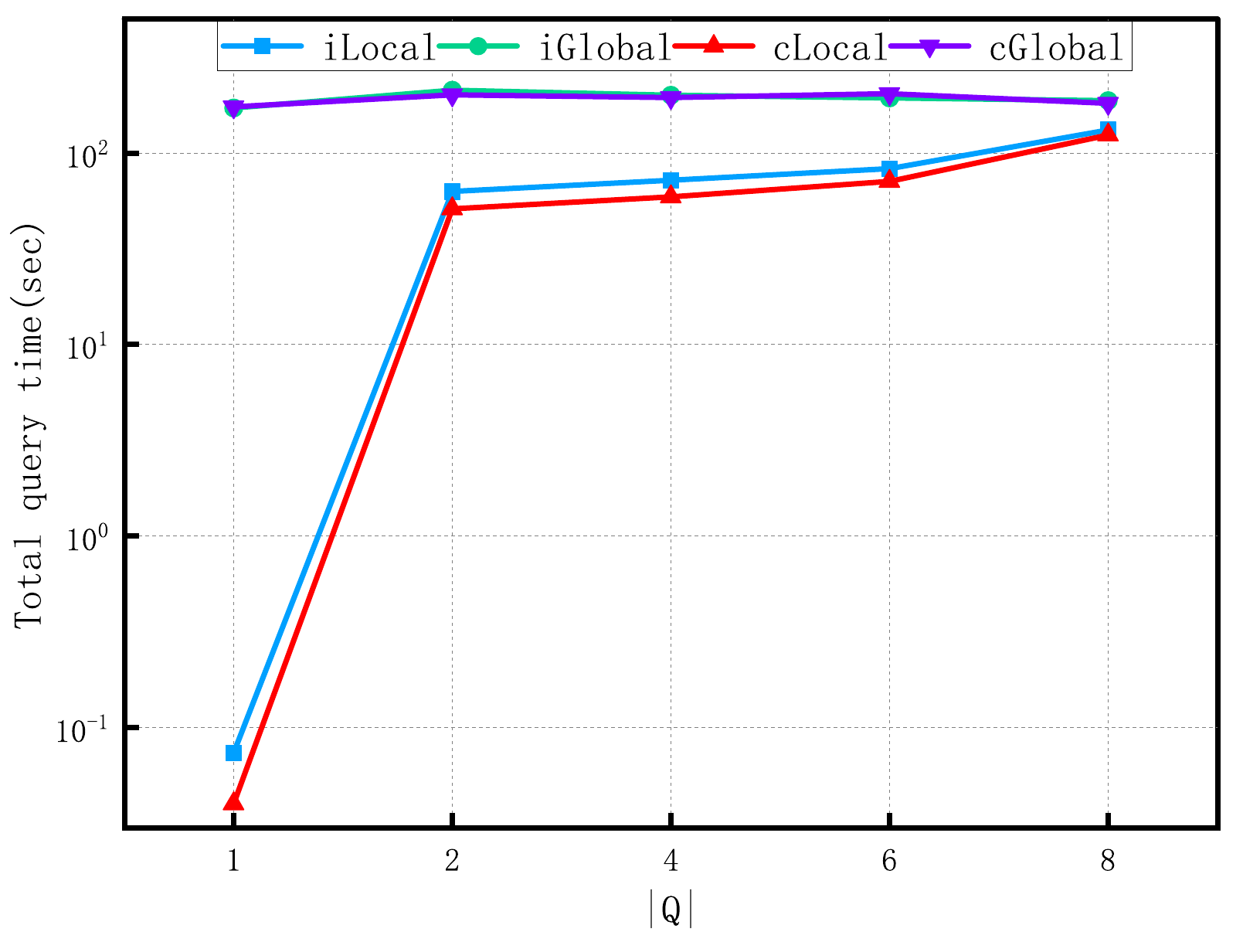}
	}%
	\subfloat[Pokec]{
		\centering
		\includegraphics[width=1.55in]{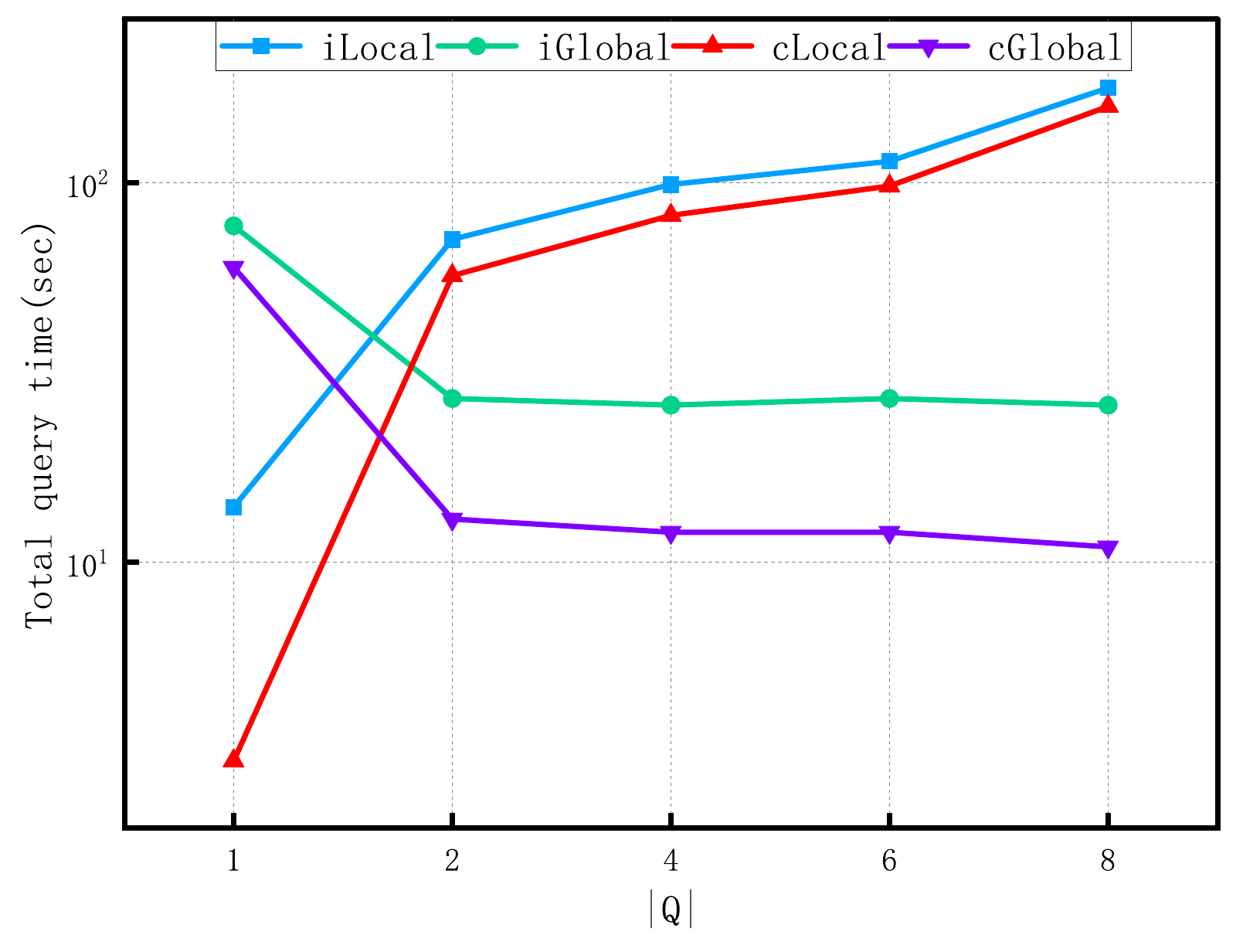}
	}%
	\centering
	\caption{CS performance for different number of $|Q|$}
\end{figure*}

In this part of the experiments,  we tested the performance of different methods, e.g., $iGlobal$, $iLocal$, $cGlobal$ (ConDTruss based $Global$), and $cLocal$ (ConDTruss based $Local$) on various datasets, e.g., Email, EAT, Slash, Pokec, and Twitter networks.

Exp-1: Changing the Degree of Query Vertices. Within real-world networks, nodes with varying degrees frequently engage in communities with differing levels of cohesion. To investigate this phenomenon, we initially arrange the vertices within each graph in descending order of degree and distribute them evenly into five equal groups. For instance, the first group comprises the top 20 percent of nodes regarding degree. In the context of each experiment set, we execute 100 queries and subsequently compute the average execution time.

As depicted in Fig 5,  the ConDTruss-based $Local$ method performs best on each dataset because ConDTruss saves many computing resources in retrieving the largest D-truss. At the same time, we can find that $cGlobal$ and $iGlobal$ are inefficient in querying the D-truss community because much time is spent in the iterative process of the Global algorithm. Across the four datasets, the efficiency of each algorithm demonstrates a tendency to remain consistent, irrespective of variations in query node degrees, which shows that the efficiency of the D-truss CS is less affected by the query node degree.

Exp-2: Changing $k_c$ and $k_f$. In this experiment, we examine the query time for CS in different datasets by changing the parameter $k_c$ or $k_f$ to examine the impact of different $k_c$ or $k_f$ values on the query time in different methods.

Figs 6(a) and 6(a) illustrate the results of changing $k_c$ and $k_f$ from 0 to 3 on EAT. Figs 6(b), (c), (d) and Figs 7(b), (c), (d) illustrate the results of changing  $k_c$ and $k_f$ from 0 to 10 on the Slash, Twitter, and Pokec, respectively. When $k_c$ or $k_f$ is increased, the run time of the four methods is decreased on all four datasets. It is because when $k_c$ or $k_f$ increase, the number of vertices and edges belonging to the returned communities are reduced, and the TC of our algorithm is determined solely by the size of the ($k_c$, $k_f$)-truss communities.

Exp-3: Changing $|Q|$. In this experiment, we test the effect of changing the size of the query node set on query efficiency. As show in Fig 8, on most datasets, the query time of each algorithm decreases as the node set increases. Among these algorithms, $eLocal$ performs the best.

\section{CONCLUSIONS}
This paper improves the efficiency of retrieving the M-D-truss by building a summarized graph index, thereby accelerating the search for the D-truss community. Initially, we propose a novel connected relation, D-truss-connected. Next, we develop the summarized graph index, ConDTruss, based on D-truss-connected, which preserves the D-truss information of the original graph. Finally, we develop a M-D-truss query algorithm based on ConDTruss. We executed comprehensive experiments on large directed graphs, and the experimental results proved that our method saves many computing resources in the retrieval process of the M-D-truss and significantly improves the efficiency of the D-truss community search.

\section*{Acknowledgments}
This work is supported by National Natural Science Foundation of China (Grant No. 69189338), Excellent Young Scholars of Hunan Province of China (Grant No. 20B625), and Changsha Natural Science Foundation (Grant No. kq2202294).

\bibliographystyle{plain}
\bibliography{reference.bib}

\end{document}